\documentclass[%
 aip,
 amsmath,amssymb,
 reprint,%
 onecolumn,
 nofootinbib
]{revtex4-2}

\renewcommand*{\thefootnote}{\fnsymbol{footnote}}

\usepackage{graphicx}
\usepackage{dcolumn}
\usepackage{bm}

\usepackage[utf8]{inputenc}
\usepackage[T1]{fontenc}
\usepackage{mathptmx}
\usepackage{etoolbox}

\usepackage{tikz}

\makeatletter
\def\@email#1#2{%
 \endgroup
 \patchcmd{\titleblock@produce}
  {\frontmatter@RRAPformat}
  {\frontmatter@RRAPformat{\produce@RRAP{*#1\href{mailto:#2}{#2}}}\frontmatter@RRAPformat}
  {}{}
}%
\makeatother

\usepackage{amsthm}

\theoremstyle{plain}
\newtheorem{theorem}{Theorem}[section]
\newtheorem{lemma}[theorem]{Lemma}
\newtheorem{corollary}[theorem]{Corollary}
\newtheorem{proposition}[theorem]{Proposition}

\theoremstyle{definition}
\newtheorem{definition}[theorem]{Definition}
\newtheorem{remark}[theorem]{Remark}

\newtheorem{assumption}[theorem]{Assumption}

\newcommand{\sub}[1]{_{\mathrm{#1}}}
\newcommand{\su}[1]{^{\mathrm{#1}}}

\usepackage{dsfont}
\newcommand{\Id}{\mathds{1}}
\newcommand{\eu}{\mathrm{e}}
\newcommand{\iu}{\mathrm{i}}
\newcommand{\di}{\mathrm{d}}

\newcommand{\N}{\mathbb{N}}
\newcommand{\Z}{\mathbb{Z}}

\newcommand{\R}{\mathbb{R}}
\newcommand{\C}{\mathbb{C}}

\newcommand{\Hi}{\mathcal{H}}
\newcommand{\F}{\mathcal{F}}

\newcommand{\norm}[1]{\left\| #1 \right\|}
\newcommand{\scal}[2]{\left\langle #1, #2 \right\rangle}

\newcommand{\set}[1]{\left\{ #1 \right\}} 

\DeclareMathOperator{\tr}{tr}
\DeclareMathOperator{\Ran}{Ran}
\DeclareMathOperator{\re}{Re}
\DeclareMathOperator{\im}{Im}
\DeclareMathOperator{\wn}{wn}

\newcommand{\PHS}{S\sub{PH}}
\newcommand{\CS}{S\sub{C}}
\newcommand{\phs}{C}
\newcommand{\cs}{S}

\begin{document}

\title[$\Z_2$ invariant for CS and PHS topological chains]{A $\Z_2$ invariant for chiral and particle-hole symmetric topological chains}
\author{Domenico Monaco}%
 \homepage{Email: \href{mailto:domenico.monaco@uniroma1.it}{domenico.monaco@uniroma1.it}}
\author{Gabriele Peluso}
 \homepage{Email: \href{mailto:gabriele.peluso@uniroma1.it}{gabriele.peluso@uniroma1.it}}
 \affiliation{Dipartimento di Matematica, ``Sapienza'' Universit\`{a} di Roma\\ Piazzale Aldo Moro 5, 00185 Rome, Italy}

\date{\today}

\begin{abstract}
We define a $\Z_2$-valued topological and gauge invariant associated to any $1$-dimensional, translation-invariant topological insulator which satisfies either particle-hole symmetry or chiral symmetry. The invariant can be computed from the Berry phase associated to a suitable basis of Bloch functions which is compatible with the symmetries. We compute the invariant in the Su--Schrieffer--Heeger model for chiral symmetric insulators, and in the Kitaev model for particle-hole symmetric insulators. We show that in both cases the $\Z_2$ invariant predicts the existence of zero-energy boundary states for the corresponding truncated models.
\end{abstract}

\maketitle


\section{Introduction} \label{sec:Intro}

The field of topological insulators has attracted the attention of both the physics and mathematics community, due to their potential applications for solid state devices and quantum computation as well as their rich mathematical structure. A topological insulator is a solid described by a Hamilton operator, acting on an appropriate Hilbert space $\mathcal{H}$, whose spectrum is gapped; this allows to define the occupied energy levels as those below the spectral gap. For translation invariant systems, the classification of topological phases of matter is then tantamount to the classification of the vector bundles over momentum space spanned by these energy states; usually, their description is mediated by numerical topological invariants, which label the different phases. In particular, different scenarios arise if the system is constrained by certain pseudo-symmetries, which can enrich or trivialize the topology of the manifold of occupied states. From the physical point of view, the most interesting feature of these materials is that, when they are truncated, a non-trivial topology for the infinite system drives the existence of metallic (i.e.\ gapless) states which are spatially localized near the cut: this statement is the celebrated principle of \emph{bulk-boundary correspondence} for topological insulators. For extended reviews of this rich research line, the reader is referred to Refs.~\onlinecite{HasanKane2010, Ando2013, Chiu_et_al2016}.

In this contribution, we will be concerned with systems which are $1$-dimensional, display a discrete translational symmetry modelling the crystalline nature of the solid at hand, and where one of two further special types of (pseudo-)symmetries, namely \emph{particle-hole} or \emph{chiral} symmetry, is present. We adopt throughout the established terminology of ``symmetries'' when referring to the particle-hole or chiral operators, even if they anti-commute rather than commute with the Hamiltonian (see below) \cite{Ryu_et_al2010}; other points of view have been recently advocated \cite{Zirnbauer2021}. The Hamiltonian of the quantum system acts on the one-particle Hilbert space of a $1$-dimensional crystal (see below) and is assumed to be spectrally gapped (without loss of generality, with a spectral gap around zero energy), in order for the system to be classified as an insulator. The crystalline structure of configuration space is reflected at the level of the Hamiltonian by requiring that there is a family of translation operators $T_\lambda$, with $\lambda$ in a lattice $\Gamma \subset \mathbb{R}$, such that $[H,T_\lambda] =0$ for all $\lambda \in \Gamma$; for simplicity we will assume $\Gamma = \Z$ in what follows. By passing to the crystal-momentum representation via the Bloch--Floquet transform (see Section~\ref{sec:BF}), the Hamiltonian can then be fibered into a family of matrices $H(k)$, with $k \in \R / (2\pi\Z) \simeq S^1$ denoting the Bloch momentum spanning the Brillouin torus. The associated spectral projection $P_-(k)$ onto the negative energy levels (the ``particles''), complemented by the orthogonal projection $P_+(k) = \Id - P_-(k)$ onto positive energy levels (the ``holes''), defines then a \emph{vector bundle} over $S^1$, called the Bloch bundle \cite{Panati2008, MonacoPanati2015}. It is the geometry of this bundle (or rather couple of bundles) that will be investigated in this paper; the reader should notice that, at this stage, both $P_-(k)$ and $P_+(k)$ define \emph{topologically trivial} bundles, as Hermitian bundles over $S^1$.

Further symmetries of the Hamiltonian are specified by the existence of a unitary or anti-unitary operator $S$ (in the case of chiral or particle-hole symmetries, respectively) such that $S\, H = -H\, S$. If, as is common in physical applications, one also requires that $S$ acts only on internal degrees of freedom in the unit cell of the lattice, then these symmetries descend also to the Bloch--Floquet fibers $H(k)$ and $P(k)$ in a $k$-independent way. These further symmetries will allow us to detect a finer geometry in the ``particle'' and ``hole'' bundles, which will be quantified in the form of a \emph{Berry phase}. We will show that this phase is quantized to an integer, and that its parity is a gauge invariant of the bundle. This will lead us to the definition of a $\Z_2$-valued gauge invariant of any chiral- or particle-hole-symmetric topological insulator, see Section~\ref{sec:Berry} (in particular Definition~\ref{dfn:invariant}). 

We will then compute this invariant in two prototypical models for these symmetry classes of topological insulators: the chiral-symmetric SSH model \cite{SSH1980}, presented in Section~\ref{sec:SSH}, and the particle-hole-symmetric Kitaev chain \cite{Kitaev2001}, discussed in Section~\ref{sec:Kitaev}. In both models, we will also show how a non-trivial invariant predicts the presence of \emph{zero-energy boundary states} when the $1$-dimensional system is truncated to a half-lattice, in agreement with bulk-boundary correspondence: in particular, in the Kitaev model, these boundary states are predicted to behave as Majorana fermions, and would represent topologically robust states amenable for applications in quantum computation. We do not claim to have a general proof of the bulk-boundary correspondence for our formulation of the bulk topological invariant, which at any rate can at most predict the parity of the number of zero-energy boundary states, and is not expected to yield a complete classification for chiral classes of topological insulators \cite{ProdanSchulz-Baldes2016, GontierMonacoPerrinRoussel2022}. Indeed, the case of chiral-symmetric topological chains and their bulk-edge correspondence, formulated in terms of a $\Z$-valued (rather than $\Z_2$-valued) topological invariant, has been already mathematically investigated, even in models where disorder breaks translation invariance \cite{ProdanSchulz-Baldes2016, GrafShapiro2018, Shapiro2020}. Nonetheless the evidence provided by these two paradigmatic models prompt us to the study of the link between our invariant and boundary states, which will be investigated in future work.


\section{Fiber decomposition and spectral projections} \label{sec:BF}

\subsection{Mathematical setting}

We now pass to a more precise formulation of the models trated in this paper. We will discuss only so-called \emph{tight-binding models}, and assume therefore that the configuration space of the quantum system is modelled after a $1$-dimensional lattice, say~$\Z$. Allowing for $N$ further internal degrees of freedom per unit cell, possibly including a sub-lattice index, the one-particle Hilbert space will be therefore taken to be
\begin{equation} \label{eqn:Hi}
\Hi = \ell^2(\Z) \otimes \C^N.
\end{equation}
The restriction of our treatment to these models, which are nonetheless of common use in condensed matter physics, is justified as follows: as will be further detailed in Remark~\ref{remark:not_zero}, the presence of a chiral or particle-hole symmetry forces the spectrum of the Hamiltonian $H$ to be symmetric around zero, namely $\mu \in \sigma(H) \Rightarrow -\mu \in \sigma (H)$. Since it is customary in quantum mechanics to assume a stability condition on the energy operator, namely that $H$ is bounded from below, the previous condition implies that the Hamiltonian is bounded, and therefore precludes the use of continuum Schr\"{o}dinger-type operators on $L^2(\R)$; notice however that there have been however recent proposals \cite{ShapiroWeinstein2022} to obtain discrete models for chiral chains starting from Schr\"odinger operators on the line, under a suitable limit of deep potential wells. We therefore also exclude Dirac-type Hamiltonians (which can be particle-hole symmetric and unbounded both from above and from below) from our treatment. One could at any rate view the discrete models introduced below also as effective models, derived from one in the continuum via projection onto a gapped spectral subspace, under an adiabatic decoupling, consisting of some ``conductance'' and ``valence'' bands of the continuum model: provided these bands exhibit this chiral or particle-hole symmetry, the tight-binding approximation would yield a symmetric lattice model.

The first set of hypotheses on the model Hamiltonian deals with the crystalline structure of configuration space. Let us therefore denote by
\[ T_\lambda \colon \Hi \to \Hi, \quad (T_\lambda \psi)_n := \psi_{n+\lambda}, \quad \psi = (\psi_n) \in \ell^2(\Z) \otimes \C^N \]
the translation operators.

\begin{assumption}[Translation-invariant insulator] \label{hyp_main}
The Hamiltonian $H$ acts on the Hilbert space $\Hi$ as in~\eqref{eqn:Hi}, and satisfies the following properties.
\begin{enumerate}
    \item It is a bounded, self-adjoint operator on $\Hi$.
    \item It is \emph{translation-invariant}, namely $[T_\lambda , H] = 0$ for all $\lambda \in \Z$.
    \item The spectrum of $H$ has a gap around zero, namely there exists $g>0$ such that $(-g,g) \cap \sigma(H) = \emptyset$.
\end{enumerate}
\end{assumption}

It is easy to verify that the action of a translation-invariant self-adjoint operator $H$ onto a vector $\psi = (\psi_n) \in \ell^2(\Z) \otimes \C^N$ is necessarily of the form
\begin{equation} \label{eqn:H->A}
(H \, \psi)_n = A_0 \, \psi_n + \sum_{j \in \N^*} A_j \, \psi_{n+j} + A_j^* \, \psi_{n-j}  
\end{equation}
for an appropriate sequence of matrices $A_j \in M_N(\C)$, $j \in \N$, with $A_0 = A_0^*$. Alternatively, if we denote by $\set{e_n}_{n \in \Z}$ the canonical basis of $\ell^2(\Z)$ whose components are $(e_n)_m = \delta_{n,\,m}$ for $n,m \in \Z$, we can write
\begin{equation} \label{eqn:Henv}
H \, (e_n \otimes v) = e_n \otimes A_0 v + \sum_{j \in \N^*} e_{n-j} \otimes A_j \, v + e_{n+j} \otimes A_j^* \, v, \quad n \in \Z, \: v \in \C^N\,,  
\end{equation}
and extend the definition of $H$ from there by linearity. 
From a physical point of view $A_j$ can be interpreted as a matrix-valued ``hopping probability'' for a particle to make a jump of $j$ sites in the lattice. Boundedness of $H$ implies some form of decay in the matrix norms $\norm{A_j}$: to simplify the setting even further, we will be content to treat \emph{finite-range operators}, and assume that

\begin{assumption}[Finite-range condition] \label{finite_range}
With $A_j \in M_N(\C)$ as above, there exists $R \in \N$ such that 
\[ A_j = 0 \quad \text{for} \quad j > R. \]
\end{assumption}

Such operators are sometimes called \emph{band Jacobi operators} in the mathematics literature~\cite{Teschl2000}. Assumptions~\ref{hyp_main} and~\ref{finite_range} will be our standing hypotheses throughout this paper, and will not be recalled further.

Finally, the models of interest for this contribution satisfy either particle-hole or chiral symmetry, as specified in the following two Assumptions.

\begin{assumption}[Particle-Hole Symmetry, PHS] \label{hyp_PHS}
There exists an \emph{anti-unitary} operator $\PHS$ on $\Hi = \ell^2(\Z) \otimes \C^N$ such that 
\begin{enumerate}
 \item $\PHS = \Id_{\ell^2(\Z)} \otimes \phs$, where $\phs$ is a anti-unitary matrix on $\C^N$ (namely, $\phs = U \, K$, where $U \in M_N(\C)$ is a unitary matrix and $K$ is a complex-conjugation operator on $\C^N$);
 \item $\PHS \, H = - H \, \PHS$.
\end{enumerate}
\end{assumption}

\begin{assumption}[Chiral Symmetry, CS] \label{hyp_CS}
There exists a \emph{unitary} operator $\CS$ on $\Hi = \ell^2(\Z) \otimes \C^N$ such that 
\begin{enumerate}
 \item $\CS = \Id_{\ell^2(\Z)} \otimes \cs$, where $\cs$ is a unitary matrix on $\C^N$;
 \item $\CS \, H = - H \, \CS$.
\end{enumerate}
\end{assumption}

A few comments on the definition of these symmetry operators are in order.
\begin{itemize}
 \item Contrary to what is usually assumed in the literature on topological phases of matter \cite{Ryu_et_al2010}, we do not need to assume anything concerning the squares of the symmetry operators $\PHS$ and $\CS$. In view of representation-theoretic reasons, it is always the case that $\CS^2 = \Id_{\Hi}$ while either $\PHS^2 = + \Id_{\Hi}$ (\emph{even} PHS) or $\PHS^2 = - \Id_{\Hi}$ (\emph{odd} PHS). These specifications will be immaterial for our discussion.
 \item The structure required on both symmetry operators, namely that they act trivially on the $\ell^2(\Z)$-leg of the tensor product defining $\Hi$ in~\eqref{eqn:Hi}, may seem restrictive, but it is often satisfied in physical models of interest such as the ones presented in Sections~\ref{sec:SSH} and~\ref{sec:Kitaev}. Notice that they imply in particular that also the symmetry operators are translation-invariant, namely $[S_*,T_\lambda]=0$ for all $\lambda \in \Z$ and for $S_* \in \set{\PHS, \CS}$.
\end{itemize}

\subsection{Bloch--Floquet representation}

It is natural and convenient to exploit the translational symmetry of the operators considered in order to simplify their analysis. This can be done by using the theory of decomposable operators, see Ref.~\onlinecite[Sec.~XIII.16]{ReedSimon1978}, and a vector-valued form of the Fourier series, called \emph{Bloch--Floquet transform} in the context of condensed matter physics. We review this method in this Section for the finite-range, translation-invariant Hamiltonians described previously, commenting in particular on the interplay of the Bloch--Floquet transform with PHS and CS. 

Denote as before by $\set{e_n}_{n \in \Z}$ the canonical basis of $\ell^2(\Z)$. Set
\[ \F \colon \ell^2(\mathbb{Z}) \otimes \mathbb{C}^N \to L^2 (S^1;\C^N), \quad \mathcal{F} ( e_n \otimes v ) := \frac{\eu^{\iu n k}}{ \sqrt{2\pi}} \, v, \qquad n \in \Z, \: v \in \C^N. \]
This is clearly a bijective linear isometry because it sends an orthonormal basis to another orthonormal basis: $\F$ is then called the Bloch--Floquet (BF) transform. We then define
\begin{equation}\label{eq:change_H}
\widehat{H} := \mathcal{F} \, H \, \mathcal{F}^{-1}\,. 
\end{equation}
Analogously, if the PHS or CS Assumptions~\ref{hyp_PHS} and~\ref{hyp_CS} hold, we can similarly define
\[
\widehat{\PHS} = \mathcal{F} \, \PHS \, \mathcal{F}^{-1} \quad \text{and} \quad \widehat{\CS} = \mathcal{F} \, \CS \, \mathcal{F}^{-1}\,.
\]
Assumptions~\ref{hyp_PHS} and~\ref{hyp_CS} then imply at once that
\begin{equation} \label{eqn:hat_sym}
\widehat{\PHS} \, \widehat{H}  = - \widehat{H} \, \widehat{\PHS} \quad \text{or} \quad \widehat{\CS} \, \widehat{H}  = - \widehat{H} \, \widehat{\CS}\,.
\end{equation}

\begin{lemma}\label{lem:decomposition_H}
The operator $\widehat{H}$ is decomposable in the sense of Ref.~\onlinecite[Sec.~XIII.16]{ReedSimon1978}, namely there exists a family of self-adjoint matrices $H(k) \in M_N(\C)$, called the \emph{fiber operators}, such that
\begin{equation} \label{eqn:decomposable}
\left(\widehat{H} u \right)(k) = H(k) \, u(k) \quad \text{for } \quad u \in L^2(S^1;\C^N)\,.
\end{equation}
With $\{A_j\}_{j \in \N}$ as in~\eqref{eqn:H->A} or~\eqref{eqn:Henv}, the fiber operators are explicitly written in the following way:
\begin{equation} \label{eqn:H(k)}
    H(k) = A_0 + \sum_{j=1}^{R} \eu^{-\iu j k} A_j + \eu^{\iu j k} A_j^* \,.
\end{equation}
\end{lemma}
\begin{proof}
In the following, we identify for notational convenience
\[ L^2 (S^1;\C^N) \simeq L^2(S^1) \otimes \C^N\,. \] 
For $n \in \Z$ and $v \in \C^N$, let us then compute, using~\eqref{eqn:Henv},
\begin{align*}
\widehat{H} \left( \frac{\eu^{\iu n k}}{\sqrt{2\pi}} \otimes v \right) & = \mathcal{F} \left[ e_n \otimes A_0 \, v + \sum_{j=1}^{R}  e_{n-j} \otimes A_j \, v + e_{n+j} \otimes A_j^* \, v \right] \\
& = \frac{\eu^{\iu n k}}{\sqrt{2 \pi}} \otimes A_0 v +  \sum_{j=1}^{R} \frac{\eu^{\iu (n-j) k}}{\sqrt{2 \pi}} \otimes A_j \, v  + \frac{ \eu^{\iu (n+j) k} }{ \sqrt{2 \pi} } \otimes A_j^* \, v \\
& = \left[A_0 + \sum_{j=1}^{R} \eu^{-\iu j k} \, A_j + \eu^{\iu j k} \, A_j^*\right] \, \left(\frac{\eu^{\iu n k }}{\sqrt{2\pi}} \otimes v \right)\,.
\end{align*}
Since the functions $\set{\eu^{\iu n k}/\sqrt{2\pi}}_{n \in \Z}$ span $L^2(S^1)$ orthonormally, the conclusion follows by linearity.
\end{proof}

\begin{remark} \label{rmk:analytic}
The matrices $H(k)$ are $2\pi$-periodic and \emph{analytic} (in fact, entire) in the variable $k$, as each entry is a finite sum of complex exponential functions. Provided $H$ is a bounded operator and the matrix norms of the $A_j$'s decay sufficiently rapidly, the same holds true even if we relax the finite-range condition, Assumption~\ref{finite_range}, by the Vitali--Porter theorem~\cite{Schiff1993}  (possibly restricting the analyticity domain to a strip of finite width around the real axis).
\end{remark}

\begin{lemma}
If the PHS Assumption~\ref{hyp_PHS} or the CS Assumption~\ref{hyp_CS} holds, then 
\[ \left(\widehat{\PHS} \, u\right) (k) = \phs \, u(-k) \quad \text{or} \quad \left(\widehat{\CS} \, u \right) (k) = \cs \, u(k), \qquad u \in L^2(S^1;\C^N), \]
where the (anti-)unitary operators $\phs, \cs \colon \C^N \to \C^N$ are as in the aforementioned Assumptions and act point-wise on $u \in L^2(S^1;\C^N)$.
\end{lemma}
\begin{proof}
As in the previous proof, by (anti-)linearity it suffices to compute the action of $\widehat{\PHS}$ and $\widehat{\CS}$ on functions $u(k)$ of the form $u(k) = \eu^{\iu n k}/\sqrt{2\pi} \otimes v$, for $n \in \Z$ and $v \in \C^N$. We have then
\[
\widehat{\PHS}\,\left( \frac{\eu^{\iu n k}}{\sqrt{2\pi}} \otimes v \right) = \mathcal{F} \left[ \PHS (e_n \otimes v ) \right] = \mathcal{F} ( e_n \otimes \phs \, v) 
 = \frac{\eu^{\iu n k}}{\sqrt{2\pi}} \otimes \phs \, v = \phs \, \left(\frac{\eu^{-\iu n k}}{\sqrt{2\pi}} \otimes v \right)
\]
as $\phs$ is \emph{anti-}linear.

Similarly we have
\[
\widehat{\CS}\,\left( \frac{\eu^{\iu n k}}{\sqrt{2\pi}} \otimes v \right) = \mathcal{F} \left[ \CS (e_n \otimes v ) \right] = \mathcal{F} ( e_n \otimes \cs \, v) 
= \frac{\eu^{\iu n k}}{\sqrt{2\pi}} \otimes \cs \, v = \cs \, \left(\frac{\eu^{\iu n k}}{\sqrt{2\pi}} \otimes v \right)
\]
as instead $\cs$ is linear.
\end{proof}

\begin{corollary} \label{cor:H(k)_sym}
If the PHS Assumption~\ref{hyp_PHS} holds, then the fiber operator $H(k)$ is anti-unitarily intertwined to the fiber operator $H(-k)$ by the relation
\[
\phs \, H(-k)= - H(k) \, \phs.
\]
Similarly, if the CS Assumption~\ref{hyp_CS} holds, then the fiber operator $H(k)$ is unitarily intertwined to the fiber operator $H(k)$ by the relation
\[
\cs \, H(k)= - H(k) \, \cs.
\]
\end{corollary}
\begin{proof}
Both statements immediately follow by combining~\eqref{eqn:hat_sym} and~\eqref{eqn:decomposable}.
\end{proof}

\begin{remark}\label{remark:not_zero}
The above condition gives us useful information regarding the spectrum of the fiber operators $H(k)$ and therefore of the Hamiltonian $H$. Indeed, assume that $v \in \C^N$ is an eigenvector of $H(k)$ of eigenvalue $E$. Let the PHS Assumption~\ref{hyp_PHS} hold. Then
\[
H(-k) \, \phs v = - \phs \, H(k) v = - \phs \,  ( E v ) = - \overline{E} \, \phs v.
\]
Due to the self-adjointness of $H(k)$, we have $E = \overline{E}$ and therefore $\phs v$ is an eigenvector of $H(-k)$ with opposite eigenvalue $-E$. Similarly, if instead the CS Assumption~\ref{hyp_CS} holds, then
\[
H(k) \, \cs v = - \cs \, H(k) v = - \cs \,  ( E v ) = - E \, \cs v
\]
and once again we find that $\cs v$ is an eigenvector for $H(k)$ with opposite eigenvalue $-E$.

From the theory of decomposable operators presented in Ref.~\onlinecite[Theorems XIII.85-86]{ReedSimon1978}, we know that the spectrum of $\widehat{H}$ (and therefore that of $H = \F^{-1} \, \widehat{H} \, \F$) is obtained as the union of the spectra of $H(k)$ for ranging $k \in S^1$. The above considerations allows us then to deduce that $\sigma(H)$ must be symmetric around $E=0$. Moreover, in view of the insulator Assumption~\ref{hyp_main}, the spectral gap of $H$ together with the analytic dependence of $H(k)$ and of its spectrum on $k \in S^1$ (see Remark \ref{rmk:analytic}) imply that $0 \notin \sigma(H(k))$ for all $k \in S^1$. Thefore we have a clear separation of positive and negative energies also for the fibre operators $H(k)$.

Notice that the Bloch momentum $k=0$ is fixed by the involution $k \mapsto - k$. Therefore, in view of what we just concluded, by the spectral theorem $\C^N$ has an orthonormal basis $v_1 , \cdots , v_N$ consisting of eigenvectors the self-adjoint matrix $H(0)$, where the first vectors $v_1 , \cdots, v_m$ have negative eigenvalues while the second vectors $v_{m+1}, \cdots, v_N$ have positive eigenvalues. The presence of $\phs$ or $\cs$ implies a the existence of a (anti-)linear bijection between the eigenspaces of $H(0)$ relative to negative and positive eigenvalues, and so $m = N-m$. Therefore, whenever Assumptions~\ref{hyp_PHS} or~\ref{hyp_CS} hold, it will be assumed that the dimension of the fiber space $\C^N$ is a an even integer, $N = 2m$.
\end{remark}

\subsection{Spectral projections}

The topological information of the insulator described by $H$, as we will detail later, is contained in the spectral projections of the fiber operators, which project over the vector subspace of $\C^N$ generated by its eigenvectors with negative eigenvalues. The collection of these subspaces leads to the definition of the \emph{Bloch bundle} \cite{Panati2008, MonacoPanati2015}.

Since $H(k)$ depends analytically on $k$ and is $2\pi$-periodic, one would expect that their eigenvalues and eigenvectors do the same. This is partially true (compare Ref.~\onlinecite[Theorem XIII.85-86]{ReedSimon1978}), but problems can occur if different eigenvalues of $H(k)$ become degenerate at some $k_0 \in S^1$. What remains true even in this situation is that the \emph{negative and positive eigenspaces} of $H(k)$ stay analytic and $2\pi$-periodic with respect to $k$, thanks to the spectral gap. More precisely, consider a PHS or CS Hamiltonian $H$ for which the considerations of the previous Subsection hold. Let $r > 0$ be such that $-2r$ is a lower bound for the spectrum of $H$, and therefore uniformly for the spectra of $H(k)$, $k \in S^1$. The \emph{Riesz formula} \cite{Kato1995}
\begin{equation} \label{eqn:Riesz}
P_- (k) := \frac{1}{2\pi \iu} \int_{\partial B_r (-r)} \di z \, \left( z \, \Id - H(k) \right)^{-1}, \quad k \in S^1,
\end{equation}
defines then an orthogonal projection, $P_-(k) = P_-(k)^* = P_-(k)^2$, on the subspace generated by the eigenvectors of $H(k)$ with negative eigenvalues. In the above, the integration is performed in the complex energy plane, along a circumference centered at $-r$ of radius $r$; the circle enclosed by this circumference only contains the negative eigenvalues of $H(k)$, where the resolvent function has its poles. The integral of a matrix should be understood as being performed entry-wise. By functional calculus, it is easily shown, e.g.\ in Ref.~\onlinecite[Proposition 2.1]{PanatiPisante2013}, that $P_-(k)$ is as regular as $H(k)$, and shares with it the same periodicity properties with respect to $k$. Of course, the same holds for the spectral projection 
\[ P_+(k) := \Id - P_-(k) \] 
onto the eigenspace of $H(k)$ associated to positive eigenvalues.

\begin{lemma}\label{lem:projection_sym}
If either Assumption~\ref{hyp_PHS} or~\ref{hyp_CS} hold, then
\begin{equation} \label{eqn:P_sym}
\phs \, P_+ (k) = P_- (-k) \, \phs \quad \text{or} \quad \cs \, P_+ (k) = P_- (k) \, \cs\,.
\end{equation}
\end{lemma} 
\begin{proof}
The statement follows at once from Corollary~\ref{cor:H(k)_sym} and Remark~\ref{remark:not_zero}.
\end{proof}


\section{Parallel transport and Bloch bases} \label{sec:parallel}

We want to quantify the topological content of the spectral projections of the fiber operators $H(k)$ into a numerical invariant. This will be done in the next Section, starting from a moving basis for the fiber Hilbert space $\C^N$ tailored around the spectral subspace of the fiber operators $H(k)$. We therefore set the following

\begin{definition}[Bloch basis] \label{def:basis_comp_proj}
Given a family of rank-$m$ projections $P(k)\in M_N(\C)$ which is $2\pi$-periodic and at least $C^1$ in $k$, we will call a family of vectors $\set{v_i (k)}_{i \in \set{1,\cdots,N}} \subset \C^N$ a \emph{Bloch basis associated to $P(k)$} if it satisfies the following proprieties for every  $k \in \mathbb{R}$.
\begin{itemize}
    \item The collection $\set{v_i (k)}_{i \in \set{1,\cdots,N}}$ forms an orthonormal basis of $\C^N$.
    \item The vectors $\set{v_i(k)}_{i \in \set{1,\ldots,m}}$ span $\Ran P(k)$; consequently, the vectors $\set{v_i(k)}_{i \in \set{m+1,\ldots,N}}$ span $\ker P(k) = \Ran [\Id - P(k)]$.
    \item Every $v_i(k)$ is $2\pi$-periodic and as regular as $P(k)$ as a function of $k \in \R$.
\end{itemize}
\end{definition}

\begin{definition}[Symmetric Bloch basis] \label{def:basis_comp_sym}
Let $P_-(k)$ be as in~\eqref{eqn:Riesz} and $P_+(k) = \Id - P_-(k)$. Assume that either Assumption~\ref{hyp_PHS} or~\ref{hyp_CS} holds, so that $N=2m$, $P_-(k)$ and $P_+(k)$ both have rank $m$, and~\eqref{eqn:P_sym} holds as well. Then a Bloch basis $\set{v_i (k)}_{i \in \set{1,\cdots,N}}$ associated to $P_-(k)$ will be called \emph{particle-hole symmetric} if
\[
\phs \, v_i(k) = v_{N-i+1}(-k) \quad \text{for all } i \in \set{1,\ldots,m}\,.
\]
Similarly, the Bloch basis will be called \emph{chiral symmetric} if
\[
\cs \, v_i(k) = v_{N-i+1}(k) \quad \text{for all } i \in \set{1,\ldots,m}\,.
\]
\end{definition}

\subsection{Parallel transport}

For $1$-dimensional systems, such as the ones at hand, it is possible to exhibit Bloch bases associated to the eigenprojections $P_-(k)$ via a tool from differential geometry called \emph{parallel transport}. We will review this construction here, and show in particular how parallel transport behaves under PHS or CS.

\begin{theorem}\label{thm:parallel_transport_existance}
If $P(k)\in M_N (\C)$ is a family of projections which is at least $C^1$ in $k$
, then there exists a unique family of operators $T(k) \in M_N(\C)$ that solves the Cauchy problem
\begin{equation}\label{eq:par_trans}
\begin{cases}
T'(k) = \left[P'(k), P(k)\right] \, T(k),\\
T(0) = \Id\,.
\end{cases}
\end{equation}
Here and in the following, the ``prime'' symbol $'$ denotes differentiation with respect to $k$. The family of operators $T(k)$ is at least as regular as $P(k)$ as a function of $k$. It further satisfies the following properties:
\begin{enumerate}
 \item \emph{Unitarity}: Each $T(k)$ is unitary, $T(k)^* = T(k)^{-1}$. 
 \item \emph{Intertwining property}: 
\begin{equation} \label{eqn:intertwining}
P(k) \, T(k) = T(k) \, P(0) \,, \quad k \in \R\,.
\end{equation}
\end{enumerate}

If $P(k)$ is also $2\pi$-periodic with respect to $k$, then $T(k)$ satisfies further the
\begin{enumerate}
\setcounter{enumi}{2}
 \item \emph{Telescopic property}: 
 \begin{equation} \label{eqn:telescopic}
  T(k + 2\pi n) = T(k) \, T(2\pi)^n\,, \quad k \in \R, \quad n \in \Z\,.
 \end{equation}
\end{enumerate}

W
e will call the family of unitaries $T(k)$ the \emph{parallel transport operators} associated to $P(k)$.
\end{theorem}
\begin{proof}
Existence, uniqueness and regularity of the solution to the Cauchy problem~\eqref{eq:par_trans} for every initial value $T_0$ follows from the standard theory of first-order linear differential equations (see e.g.\ Ref.~\onlinecite[Section IV.14.VI]{Walter1998}). 

\paragraph{\emph{Unitarity}} 
To prove that $T(k)$ is unitary, it suffices to notice that the (non-autonomous) generator of the differential equation is skew-adjoint:
\[
\left[P'(k),P(k)\right]^* = -\left[P'(k),P(k)\right]\,.
\]

\paragraph{\emph{Intertwining property}}
The intertwining property~\eqref{eqn:intertwining} clearly holds at $k=0$
. To conclude that it holds for all $k \in \R$, one computes $[T(k) ^* \, P(k) \, T(k)]'$ and shows that it vanishes. We have that
\[
\frac{\di}{\di k} \, T(k) ^* \, P(k) \, T(k) 
 = T(k)^* \, \left\{ - \left[P'(k),P(k)\right] \, P(k) + P'(k) + P(k) \, \left[P'(k),P(k)\right] \right\} \, T(k)\,.
\]
The term in curly brackets vanishes, as one can verify with a straightforward algebraic computation using the identity $P(k) \, P'(k) \, P(k) = 0$ (following from differentiation of the relation $P(k) = P(k)^2$). 

\paragraph{\emph{Telescopic property}}
To prove~\eqref{eqn:telescopic}, let us define $A(k) := T(k+2\pi n)$ for fixed $n \in \Z$. 
By direct inspection, $A(k)$ solves the Cauchy problem~\eqref{eq:par_trans} with $A(0) = T(2\pi n)$, as the family of projections $P(k)$ is $2\pi$-periodic. 
The same is true for $B(k) := T(k) \, T(2\pi n)$, and by uniqueness it must hold that
\[ T(k+2\pi n) = T(k) \, T(2\pi n) \quad \text{for all } k \in \R, \; n \in \Z \,. \]
The above implies in particular
\[ T(2\pi n) = T(2\pi)^n, \quad n \in \Z \]
from which~\eqref{eqn:telescopic} follows.
\end{proof}

\begin{remark}\label{obs:parallel_transport_+-}
Uniqueness of the solution of the Cauchy problem in the statement of Theorem~\ref{thm:parallel_transport_existance} implies that the parallel transport operators associated to the families $P(k)$ and $Q(k) := \Id - P(k)$ coincide, since they share the same generator of the linear differential equation:
\[ \left[Q'(k), Q(k)\right] = \left[(\Id-P(k))', \Id-P(k)\right] = \left[P'(k),P(k)\right]\,. \]
\end{remark}

\begin{lemma}\label{lem:transport_symmetry}
Let $P_-(k)$ be the family of projections defined in~\eqref{eqn:Riesz}, and denote by $T(k)$ the parallel transport unitaries associated to $P_-(k)$. Assume that either Assumption~\ref{hyp_PHS} or~\ref{hyp_CS} holds. Then
\[
\phs \, T(k) = T(-k) \, \phs \quad \text{or} \quad \cs \, T(k) = T(k) \, \cs \,, \qquad k \in \R.
\]
\end{lemma}
\begin{proof}
Under Assumption~\ref{hyp_PHS}, let us denote by $U(k) := \phs^{-1} \, T(-k) \, \phs$. Compute
\begin{align*}
U'(k) & = - \phs^{-1} \, \left[P_-'(-k),P_-(-k)\right] \, T(-k) \, \phs = \left[-\phs^{-1} \, P_-'(-k) \, \phs, \phs^{-1} \, P_-(-k) \, \phs\right] \, \phs^{-1} \, T(-k) \, \phs \\
& = \left[P_+'(k) , P_+(k)\right] \, U(k)
\end{align*}
where we used Lemma~\ref{lem:projection_sym} for the last equality. Since $U(0) = \Id$ this means that $U(k)$ solves the Cauchy problem~\eqref{eq:par_trans} for the parallel transport operators associated to $P_+(k) = \Id - P_-(k)$. By Remark~\ref{obs:parallel_transport_+-} we conclude that $U(k) = T(k)$.

Arguing similarly under Assumption~\ref{hyp_CS}, one can show that $V(k) := \cs^{-1} \, T(k) \, \cs$ and $T(k)$ solve the same Cauchy problem, and therefore coincide.
\end{proof}

\subsection{Construction of symmetric Bloch bases}

Using the parallel transport unitaries $T(k)$ we can try to construct a symmetric Bloch basis. If we fix an orthonormal basis $\set{v_1(0) , \cdots, v_{2m}(0)} \subset \C^{2m}$ of eigenvectors of $H(0)$ such that the first half have negative eigenvalues while the second half have positive eigenvalues, we can define the family of vectors
\[ \tilde{v}_i(k) := T(k) \, v_i(0), \quad k \in \R. \]
These vectors form an orthonormal basis of $\C^N$ that depends smoothly on $k$; moreover, in view of the intertwining property~\eqref{eqn:intertwining}, we have that
\[
P_-(k) \tilde{v}_i(k) = \begin{cases}
\tilde{v}_i(k) & \text{if } i \le m \,, \\ 
0 & \text{if } i > m\,,
\end{cases} \quad \text{while} \quad
P_+(k) \tilde{v}_i(k) = \begin{cases}
0 & \text{if } i \le m\,, \\
\tilde{v}_i(k), & \mbox{if }  i > k\,.
\end{cases}
\]
Finally, if the model enjoys PHS or CS, then by Remark~\ref{remark:not_zero} we can further choose the basis of eigenvectors of $H(0)$ in such a way that
\begin{equation} \label{eqn:symm_at_0}
\phs \, v_i(0) = v_{N-i+1}(0)\,, \quad \text{or} \quad \cs \, v_i(0) = v_{N-i+1}(0)\,.
\end{equation}
Then, in view of Lemma~\ref{lem:transport_symmetry}, we conclude that for $i \in \set{1, \ldots, m}$
\[ \phs \, \tilde{v}_i(k) = \phs \, T(k) \, v_i(0) = T(-k) \, \phs \, v_i(0) = T(-k) \, v_{N-i+1}(0) = \tilde{v}_{N-i+1}(-k)\,, \]
or similarly
\[ \quad \cs \, \tilde{v}_i(k) = \tilde{v}_{N-i}(k)\,. \]
This is almost enough to conclude that $\set{\tilde{v}_i(k)}_{i \in \set{1, \ldots,N}}$ is a symmetric Bloch basis, but unfortunately in general it will not be $2\pi$-periodic since $T(2\pi) \ne T(0) = \Id$. Therefore we need to adjust this construction.

Let us then focus on the unitary matrix $T(2\pi) \in M_N(\C)$, called the \emph{holonomy unitary}. By the spectral theorem, $T(2\pi)$ can be diagonalized by a unitary transformation $V$:
\[ T(2\pi) = V^{-1} \, \begin{bmatrix} \eu^{\iu\phi_1} & \cdots & 0 \\
\vdots & \ddots & \vdots \\
0 & \cdots & \eu^{\iu\phi_N}
\end{bmatrix} \, V\,, \quad \text{with} \quad V^{-1} = V^*, \quad \phi_i \in [0,2\pi) \ \forall i \in \set{1, \ldots, N}\,. \]
With this choice of the phases for the eigenvalues of $T(2\pi)$, we let $X \in M_N(\C)$ be the self-adjoint matrix
\begin{equation} \label{eqn:X}
X = -\iu \, \ln T(2\pi) := V^{-1} \, \begin{bmatrix} \phi_1 & \cdots & 0 \\
\vdots & \ddots & \vdots \\
0 & \cdots & \phi_N
\end{bmatrix} \, V
\end{equation}
so that $T(2\pi) = \eu^{\iu X}$. It is worth to notice that $X$ and $T(2\pi)$ share the same eigenspaces.

\begin{theorem}\label{thm:basis_comptible_projection}
Let $P_-(k) \in M_N(\C)$ be the rank-$m$ projections defined in~\eqref{eqn:Riesz}. Let also $\set{v_1(0) , \cdots, v_N(0)} \subset \C^N$ be an orthonormal basis of eigenvectors of $H(0)$ such that $\set{v_1(0) , \cdots, v_{m}(0)}$ span $\Ran P_-(0)$ while $\set{v_{m+1}(0) , \cdots, v_N(0)}$ span $\ker P_-(0)$. Finally, denote by $T(k)$ the family of parallel transport unitaries associated to $P_-(k)$. Then the family of vectors
\begin{equation} \label{eqn:par_trans_v}
v_i(k) := T(k) \, \eu^{-\iu k X/2\pi} \, v_i(0)\,, \quad i \in \set{1, \ldots, N}
\end{equation}
define a Bloch basis associated to $P_-(k)$ in the sense of Definition~\ref{def:basis_comp_proj}.

If furthermore Assumption~\ref{hyp_PHS} or~\ref{hyp_CS} holds, then the vectors defined in~\eqref{eqn:par_trans_v} form a symmetric Bloch basis associated to $P_-(k)$, in the sense of Definition~\ref{def:basis_comp_sym}.
\end{theorem}
\begin{proof} We split the proof in a few steps.

\paragraph{\emph{Spanning property}}
The vectors $\set{v_i(k)}_{i \in \set{1,\ldots,N}}$ are clearly orthonormal as they are obtained from the orthonormal basis of eigenvectors of $H(0)$ by means of unitary transformations. Let us check that the first $m$ vectors span $\Ran P_-(k)$, so that the last $N-m$ will span $\Ran P_+(k)$. To this end, it suffices to notice that the intertwining property~\eqref{eqn:intertwining} of the parallel transport unitaries yields in particular
\[ P_-(2\pi) \, T(2 \pi) = P_-(0) \, T(2\pi) = T(2\pi) \, P_-(0) \]
in view of the $2\pi$-periodicity of the family of projections. So $T(2\pi)$ commutes with $P_-(0)$, and as the matrix $X$ in~\eqref{eqn:X} is obtained from $T(2\pi)$ through functional calculus, so does $X$ and therefore $\eu^{-\iu k X/2\pi}$ for all $k \in \R$. It then follows that the vectors $\eu^{-\iu k X/2\pi} \, v_i(0)$  for $i \in \set{1, \ldots, m}$ are orthonormal and in the range of $P_-(0)$, and are then mapped orthonormally by $T(k)$ to the range of $P_-(k)$ in view of the intertwining property again.

\paragraph{\emph{Regularity}}
Since $P_-(k)$ is analytic in $k$, $T(k)$ is as well by Theorem~\ref{thm:parallel_transport_existance}. Also the matrix
\[ \eu^{- \iu k X/2\pi} = V^{-1} \, 
\begin{bmatrix} 
\eu^{-\iu k \phi_1/2\pi} & \cdots & 0 \\
\vdots & \ddots & \vdots \\
0 & \cdots & \eu^{-\iu k \phi_N/2\pi}
\end{bmatrix} \, V \]
depends analytically on $k$. We conclude that~\eqref{eqn:par_trans_v} defines an analytic $\C^N$-valued function of $k$.

\paragraph{\emph{Periodicity}}
To prove that $v_i(k)$ is $2\pi$-periodic in $k$, we will exploit the telescopic property~\eqref{eqn:telescopic} of the parallel transport unitaries. We have indeed, for $k \in \R$ and $n \in \Z$, 
\begin{align*}
v_i(k+2\pi n) & = T(k+2\pi n) \, \eu^{-\iu (k+2\pi n) X/2\pi} \, v_i(0) = T(k) \, T(2\pi)^n \, \eu^{-\iu n X} \, \eu^{-\iu k X/2\pi} \, v_i(0) \\
& = T(k) \, \eu^{-\iu k X/2\pi} \, v_i(0) = v_i(k)\,,
\end{align*}
since by definition $T(2\pi)^n = \eu^{\iu n X}$.

\paragraph{\emph{Symmetry}}
Under the CS hypothesis, we know from Lemma~\ref{lem:transport_symmetry} that $\cs$ commutes with $T(k)$, so in particular it commutes with $T(2\pi)$ and by functional calculus with $X$ and with $\eu^{-\iu k X /2\pi}$.
Therefore it holds that
\[
\cs \, v_i (k) 
= \cs \, T(k) \, \eu^{-\iu k X/2\pi} \, v_i(0) = T(k) \, \eu^{-\iu k X/2\pi} \, \cs \, v_i(0) 
= T(k) \, \eu^{-\iu k X/2\pi} \, v_{N-i+1}(0) = v_{N-i+1} (k)
\]
as wanted.

Let us now assume the PHS hypothesis. Lemma~\ref{lem:transport_symmetry} and the telescopic property~\eqref{eqn:telescopic} together give that
\[ \phs \, T(2\pi) = T(-2\pi) \, \phs = T(2\pi)^{-1} \, \phs \quad \Longleftrightarrow \quad \phs \, \eu^{\iu X} = \eu^{-\iu X} \, \phs \]
which by functional calculus implies for all $k \in \R$
\[ \phs \, \eu^{-\iu k X/2\pi} = \eu^{\iu k X/2\pi} \, \phs\,. \]
Therefore it holds that
\[
\phs \, v_i (k) 
= \phs \, T(k) \, \eu^{-\iu k X/2\pi} \, v_i(0) = T(-k) \, \eu^{\iu k X/2\pi} \, \phs \, v_i(0) 
= T(-k) \, \eu^{\iu k X/2\pi} \, v_{N-i+1}(0) = v_{N-i+1} (-k)
\]
which concludes the proof.
\end{proof}


\section{Berry phase and the $\Z_2$ invariant} \label{sec:Berry}

We are finally in position to extract the topological invariant out of any Bloch basis associated to the spectral projections $P_-(k)$ of the fiber Hamiltonians $H(k)$, which we know exist thanks to the results of the previous Section. The invariant will be formulated starting from the (\emph{abelian}) \emph{Berry phase} of the Bloch basis, whose definition we recall below.

\begin{definition}[Berry phase]
Let $P(k) \in M_N(\C)$ be a $2\pi$-periodic family of rank-$m$ projections which is at least $C^1$ in $k$, and $\{v_i(k) \}_{i\in\set{1 , \cdots , N}}$ be a Bloch basis associated to $P(k)$ in the sense of Definition~\ref{def:basis_comp_proj}.
The \emph{Berry phase} of the Bloch basis is the following integral:
\begin{equation}\label{eq:Berry_phase}
    \frac{1}{2\pi \iu}\, \int_{S^1} \di k \, \sum_{i=1}^{N} \scal{v_i(k)}{v_i'(k)} \,.
\end{equation}
\end{definition}

The same formula, in general, gives the holonomy phase of any moving basis (i.e.\ global frame) for the trivial bundle $S^1 \times \C^N$. 

This Section will be devoted to proving some properties of the Berry phase defined above: it will be argued that it takes integer values, and that under PHS or CS the parity of this integer is a quantity which depends on the projections $P_-(k)$, and not on the choice of a Bloch basis compatible with the symmetries. This will lead to the formulation of the $\Z_2$ invariant.

\subsection{Properties of the Berry phase}

Much of the following discussion will be based on the next result regarding homotopy classes of unitary-matrix-valued maps on the circle. This statement is well-known from algebraic topology, see e.g.\ Ref.~\onlinecite[Proposition 13.11]{Hall2015}, but we provide a sketch of some aspects of the proof in Appendix~\ref{sec:appendix} for the reader's convenience. 

\begin{theorem}[Winding number of unitary-matrix-valued maps] \label{thm:p1(UN)}
Let $\pi_1(U(N))$ denote the group of (smooth) homotopy classes of (smooth) maps $S^1 \to U(N)$, with $U(N) := \set{ U \in M_N(\C) : U^* = U^{-1} }$, endowed with point-wise multiplication of maps as group operation. Then there is a group isomorphism
\begin{equation} \label{eqn:wn}
\wn \colon \pi_1(U(N)) \stackrel{\sim}{\longrightarrow} \Z, \qquad [U \colon S^1 \to U(N)] \mapsto \frac{1}{2\pi\iu} \, \int_{S^1} \di k \, \tr \big( U(k)^* \, U'(k) \big)\,.
\end{equation}
The integer $\wn([U])$ is called the \emph{winding number} of (the homotopy class $[U]$ of) the map $U \colon S^1 \to U(N)$.
\end{theorem}

We start by computing the Berry phase of the Bloch basis constructed in the previous Section.

\begin{theorem}\label{thm:global_Berry_phase}
Under the assumptions of Theorem~\ref{thm:basis_comptible_projection}, we have that
\[
\frac{1}{2\pi \iu} \, \int_{S^1} \di k \, \sum_{i=1}^{N} \scal{v_i (k)}{v_i' (k)} = - \frac{1}{2\pi} \, \tr X \quad \in  \Z \,
\] 
where $\set{v_i (k)}_{i \in \set{1, \ldots, N}}$ is as in~\eqref{eqn:par_trans_v} and $X$ is as in~\eqref{eqn:X}. In particular, the Berry phase above is independent of the choice of $\set{v_i (0)}_{i \in \set{1, \ldots, N}}$ as in the statement of Theorem~\ref{thm:basis_comptible_projection}.
\end{theorem}
\begin{proof}
With the notation of Theorem~\ref{thm:basis_comptible_projection}, let us define $W(k) := T(k) \, \eu^{-\iu k X /2\pi}$, so that $v_i(k) = W(k) \, v_i(0)$ for $i \in \set{1,\ldots,N}$. We can compute then
\begin{align*}
\frac{1}{2\pi \iu} \, \int_{S^1} \di k \, \sum_{i=1}^{N} \scal{v_i(k)}{v_i'(k)} & = \frac{1}{2\pi \iu} \, \int_{S^1} \di k \, \sum_{i=1}^{N} \scal{W(k) \, v_i(0)}{W'(k) \, v_i(0)} = \frac{1}{2\pi \iu} \, \int_{S^1} \di k \, \sum_{i=1}^{N} \scal{v_i(0)}{W(k)^* \, W'(k) \, v_i(0)} \\
& = \frac{1}{2\pi \iu} \, \int_{S^1} \di k \, \tr \big( W(k)^* \, W'(k) \big)
\end{align*}
owing to the fact that $\set{v_i(0)}_{i \in \set{1,\ldots,N}}$ is an orthonormal basis of $\C^N$. In view of Theorem~\ref{thm:p1(UN)}, this allows to conclude that
\[ \frac{1}{2\pi \iu} \, \int_{S^1} \di k \, \sum_{i=1}^{N} \scal{v_i(k)}{v_i'(k)} = \wn([W]) \in \Z\,. \]

To prove the equality claimed in the statement, we compute then
\[
W'(k) 
= \frac{\di}{\di k} \big( T(k) \, \eu^{-\iu k X/2\pi} \big) = T'(k) \, \eu^{-\iu k X/2\pi} + T(k) \, \eu^{-\iu k X/2\pi} \, \frac{X}{2\pi\iu} 
= [P_-'(k),P_-(k)] \, W(k) + W(k) \, \frac{X}{2\pi\iu}
\]
and therefore
\[ W(k)^* \, W'(k) = W(k)^* \, [P_-'(k),P_-(k)] \, W(k) + \frac{X}{2\pi\iu}\,. \]
Consequently
\[
\tr \big(W(k)^* \, W'(k) \big) 
= \tr \big( W(k)^* \, [P_-'(k),P_-(k)] \, W(k) \big) + \frac{1}{2\pi \iu} \, \tr X 
= \tr \big( [P_-'(k),P_-(k)] \big) + \frac{1}{2\pi \iu} \, \tr X
\]
due to the invariance of the trace under the unitary conjugation by $W(k)$. Now, the first summand on the right-hand side of the above vanishes because commutators are traceless, and we conclude that
\[ \wn([W]) = \frac{1}{2\pi\iu} \, \int_{S^1} \di k \, \tr \big(W(k)^* \, W'(k) \big) = \frac{1}{(2\pi\iu)^2} \, \tr X \, \int_{S^1} \di k = - \frac{1}{2\pi} \, \tr X \]
as claimed.

Finally $\tr X$ does not depend on the choice of the basis $\set{v_i(0)}_{i \in \set{1,\ldots,N}}$ for $\C^N$, so neither does the Berry phase of the Bloch basis $\set{v_i(k)}_{i \in \set{1,\ldots,N}}$.
\end{proof}

\begin{remark}\label{obs:det(T)=1}
Since $T(2\pi) = \eu^{\iu X}$ and $\tr(X) = 2 \pi n$ for $n = - \wn([W]) \in \Z$, it must be that $\det T(2\pi) = \eu^{\iu \, \tr X} = \eu^{2\pi \iu n} = 1$. So, while the parallel transport unitaries $T(k)$ are not $2\pi$-periodic in $k$ in general (compare~\eqref{eqn:telescopic}), the determinant map $k \mapsto \det T(k)$ is, and therefore has a well defined winding number in view of Proposition~\ref{cor:top_group}, which can be alternatively used to compute (up to a sign) the Berry phase of the Bloch basis in~\eqref{eqn:par_trans_v}. 
\end{remark}

The next result will tell us that, when a change of Bloch basis is performed (also called a \emph{change of Bloch gauge}), the Berry phase changes by an integer. In combination with the previous result, we can conclude that Berry phases are always integer-valued.

\begin{theorem}\label{cor:change_gauge_choice}
Let $\set{v_i (k)}_{i \in \set{1, \ldots, N}}$ be is as in~\eqref{eqn:par_trans_v} and let $\set{u_i (k)}_{i \in \set{1, \ldots, N}}$ be any Bloch basis associated to $P_-(k)$, in the sense of Definiton~\ref{def:basis_comp_proj}. Define the change-of-basis matrix $G(k)$, called \emph{Bloch gauge}, such that $u_i(k) = G(k) \, v_i(k)$ for all $i \in \set{1,\ldots,N}$. Then $G(k)$ is unitary, $2\pi$-periodic and as regular in $k$ as the Bloch bases. Moreover
\[
\frac{1}{2\pi \iu} \, \int_{S^1} \di k \, \sum_{i=1}^{N} \scal{u_i(k)}{u_i'(k)} = \frac{1}{2\pi \iu} \, \int_{S^1} \di k \, \sum_{i=1}^{N} \scal{v_i(k)}{v_i'(k)} + \wn([G]) \quad \in \Z 
\]
where $\wn([G])$ denotes the winding number of the map $G \colon S^1 \to U(N)$.
\end{theorem}
\begin{proof}
For each $k \in \R$, the matrix $G(k)$ is unitary, as it maps one orthonormal basis into another. Moreover, in the $2\pi$-periodic and regular orthonormal basis $\set{v_i (k)}_{i \in \set{1, \ldots, N}}$ for $\C^N$, the Bloch gauge $G(k)$ has entries
\[ \scal{v_j(k)}{G(k)\,v_i(k)} = \scal{v_j(k)}{u_i(k)}, \quad i,j \in \set{1, \ldots,N}\,, \]
and therefore it is itself $2\pi$-periodic and regular, as wanted.

We can then compute
\begin{align*}
\frac{1}{2\pi \iu} & \, \int_{S^1} \di k \, \sum_{i=1}^{N} \scal{u_i(k)}{u_i'(k)} = \frac{1}{2\pi \iu} \, \int_{S^1} \di k \, \sum_{i=1}^{N} \scal{G(k) \, v_i(k)}{\partial_k \big( G(k) \, v_i(k) \big)} \\
& = \frac{1}{2\pi \iu} \, \int_{S^1} \di k \, \sum_{i=1}^{N} \scal{G(k) \, v_i(k)}{G(k) \, v_i'(k)} + \frac{1}{2\pi \iu} \, \int_{S^1} \di k \, \sum_{i=1}^{N} \scal{G(k) \, v_i(k)}{G'(k) \, v_i(k)} \\
& = \frac{1}{2\pi \iu} \, \int_{S^1} \di k \, \sum_{i=1}^{N} \scal{v_i(k)}{v_i'(k)} + \frac{1}{2\pi \iu} \, \int_{S^1} \di k \, \sum_{i=1}^{N} \scal{v_i(k)}{G(k)^* \, G'(k) \, v_i(k)} \\
& = \frac{1}{2\pi \iu} \, \int_{S^1} \di k \, \sum_{i=1}^{N} \scal{v_i(k)}{v_i'(k)} + \frac{1}{2\pi \iu} \, \int_{S^1} \di k \, \tr \big( G(k)^* \, G'(k) \big) \\
& = \frac{1}{2\pi \iu} \, \int_{S^1} \di k \, \sum_{i=1}^{N} \scal{v_i(k)}{v_i'(k)} + \wn([G])\,,
\end{align*}
as claimed.
\end{proof}

\begin{remark} \label{rmk:G+-}
By definition, a Bloch basis $\set{u_i (k)}_{i \in \set{1, \ldots, N}}$ associated to $P_-(k)$ has the first $m$ vectors in $\Ran P_-(k)$ and the last $N-m$ in $\Ran P_+(k)$, where as usual $P_+(k) = \Id - P_-(k)$. Therefore, the Bloch gauge $G(k)$  which maps the Bloch basis $\set{v_i (k)}_{i \in \set{1, \ldots, N}}$ in~\eqref{eqn:par_trans_v} to $\set{u_i (k)}_{i \in \set{1, \ldots, N}}$ has a block-diagonal form in the decomposition $\C^N = \Ran P_-(k) \oplus \Ran P_+(k)$, i.e.
\begin{equation} \label{eqn:G+-}
G(k) = \begin{bmatrix} G_-(k) & 0 \\ 0 & G_+(k) \end{bmatrix}
\end{equation}
where $G_\pm(k) = P_\pm(k) \, G(k) \, P_\pm(k)$ seen as a linear operator on $\Ran P_\pm(k)$. In particular
\[ u_i(k) = \begin{cases}
G_-(k) \, v_i(k) & \text{if } i \in \set{1,\ldots,m}\,, \\
G_+(k) \, v_i(k) & \text{if } i \in \set{m+1,\ldots,N}\,.
\end{cases} \]
\end{remark}

\subsection{Symmetric Berry phases}

We now turn to the investigation of how PHS and CS affect the value of Berry phases. 

\begin{proposition} \label{prop:even}
Let Assumption~\ref{hyp_PHS} or~\ref{hyp_CS} hold. Let $\set{v_i (k)}_{i \in \set{1, \ldots, N}}$ be is as in~\eqref{eqn:par_trans_v} and let $\set{u_i (k)}_{i \in \set{1, \ldots, N}}$ be any symmetric Bloch basis associated to $P_-(k)$, in the sense of Definiton~\ref{def:basis_comp_sym}. Finally, let $G(k)$ be the Bloch gauge defined in Theorem~\ref{cor:change_gauge_choice}. Then
\[ \wn([G]) \in 2 \, \Z\,. \]
\end{proposition}
\begin{proof}
We will use the block decomposition of $G(k)$ from~\eqref{eqn:G+-}. As usual we will denote by $m$ the rank of $P_-(k)$, so that in particular $N=2m$.

\paragraph{\emph{CS}}
Under Assumption~\ref{hyp_CS}, we have that
\[ v_{N-i+1}(k) = \cs \, v_i(k) \quad \text{and} \quad u_{N-i+1}(k) = \cs \, u_i(k) \quad \text{for } i \in \set{1, \ldots, m}\,. \]
We have also, for  $i \in \set{1, \ldots, m}$,
\[ G_+(k) \, \cs \, v_i(k) = G_+(k) \, v_{N-i+1}(k) = u_{N-i+1}(k) = \cs \, u_i(k) = \cs \, G_-(k) \, v_i(k)\,. \]
Since $\set{v_i(k)}_{i \in \set{1,\ldots,m}}$ span $\Ran P_-(k)$ orthonormally, we conclude that
\[ G_+(k) \, P_+(k) \, \cs = G_+(k) \, \cs \, P_-(k)= \cs \, G_-(k) \, P_-(k)\,.\]
Since $G_\pm(k) = P_\pm(k) \, G(k) \, P_\pm(k)$ by definition, we conclude from the above that
\[ P_+(k) \, G(k) \, P_+(k) \, \cs = \cs \, P_-(k) \, G(k) \, P_-(k) \]
and therefore
\[ \det G(k) = \det G_-(k) \cdot \det G_+(k) = \big( \det G_-(k) \big)^2\,. \]
(In the above, it should be noted that $\det G$ is the determinant of a linear operator on $\C^N = \C^{2m}$, while $\det G_\pm$ are determinants of linear operators on $\Ran P_\pm(k) \simeq \C^m$.) In view of the results of Appendix~\ref{sec:appendix}, the winding number $\wn([G])$ can be computed as the winding $w(\det G)$ of the map $\det G \colon S^1 \to S^1$, and additivity of this winding number (Proposition~\ref{cor:top_group}) yields
\[ \wn([G]) = 2 \, \wn([G_-]) \in 2 \, \Z\,. \]

\paragraph{\emph{PHS}}
Under Assumption~\ref{hyp_PHS}, we have this time
\[ G_+(-k) \, \phs \, v_i(k) = G_+(-k) \, v_{N-i+1}(-k) = u_{N-i+1}(-k) = \phs \, u_i(k) = \phs \, G_-(k) \, v_i(k) \]
for $i \in \set{1, \ldots, m}$, and therefore
\[ P_+(-k) \, G(-k) \, P_+(-k) \, \phs = \phs \, P_-(k) \, G(k) \, P_-(k) \,. \]
By antilinearity of $\phs$, the above implies that 
\[ \det G_+(-k) = \overline{\det G_-(k)} = \det G_-(k)^* = \det G_-(k)^{-1} \]
and in view of Remark~\ref{rmk:wind_inv}
\[ w(\det G_+) = - w(\det G_+ \circ \iota) = - w(\det G_-^{-1}) = w(\det G_-) \]
where $\iota(k) := -k$. As before, this identity allows to deduce that $\wn([G]) = \wn([G_-]) + \wn([G_+]) = 2 \, \wn([G_-]) \in 2\,\Z$, which concludes the proof.
\end{proof}

\begin{corollary}\label{cor:invariant}
Under the assumptions of Proposition~\ref{prop:even}, the $\bmod\ 2$ equivalence class 
\[ \left[ \frac{1}{2\pi \iu} \, \int_{S^1} \di k \, \sum_{i=1}^{2m} \scal{u_i(k)}{u_i'(k)} \right] = \left[ \frac{1}{\pi \iu} \, \int_{S^1} \di k \, \sum_{i=1}^{m} \scal{u_i(k)}{u_i'(k)} \right] \quad \in \Z_2
\]
does not depend on the choice of a symmetric Bloch basis $\set{u_i(k)}_{i \in \set{1,\ldots,N}}$ associated to $P_-(k)$.
\end{corollary}
\begin{proof}
The independence of the $\bmod\ 2$ reduction of the Berry phase from the Bloch basis is a direct consequence of Theorem~\ref{cor:change_gauge_choice} and Proposition~\ref{prop:even}, so we only need to prove the rewriting of the Berry phase claimed in the statement.

\paragraph{\emph{CS}}
Under Assumption~\ref{hyp_CS} it suffices to notice that for a chiral symmetric Bloch basis $\set{u_i(k)}_{i \in\set{1,\ldots,N}}$
\[ \scal{u_{N-i+1}(k)}{u_{N-i+1}'(k)} = \scal{\cs \, u_i(k)}{\cs \, u_i'(k)} = \scal{u_i(k)}{u_i'(k)}\,, \quad i \in \set{1,\ldots,m}\,, \]
and therefore
\[ \sum_{i=1}^{2m} \scal{u_i(k)}{u_i'(k)} = 2 \, \sum_{i=1}^{m} \scal{u_i(k)}{u_i'(k)}\,. \]

\paragraph{\emph{PHS}}
This time, under Assumption~\ref{hyp_PHS}, we have that for a particle-hole symmetric Bloch basis
\[ u_{N-i+1}(-k) = \phs \, u_i(k) \quad \Longrightarrow \quad - u_{N-i+1}'(-k) = \phs \, u_i'(k)\,, \quad i \in \set{1,\ldots,m}\,, \]
and therefore
\[ \scal{u_{N-i+1}(-k)}{u_{N-i+1}'(-k)} = - \scal{\phs \, u_i(k)}{\phs \, u_i'(k)} = - \overline{\scal{u_i(k)}{u_i'(k)}}\,, \quad i \in \set{1,\ldots,m}\,, \]
due to the antilinearity of $\phs$. Now, the normalization $\scal{u_i(k)}{u_i(k)} \equiv 1$ implies that $\scal{u_i(k)}{u_i'(k)}$ is a purely imaginary number, which yields
\[ \scal{u_{N-i+1}(-k)}{u_{N-i+1}'(-k)} = \scal{u_i(k)}{u_i'(k)}\,, \quad i \in \set{1,\ldots,m}\,. \]

Notice now that, by changing variables to $\kappa = - k$, we can rewrite
\begin{align*}
\int_{0}^{2\pi} \di k \, \sum_{j=m+1}^{2m} \scal{u_j(k)}{u_j'(k)} & = - \int_{0}^{-2\pi} \di \kappa \, \sum_{j=m+1}^{2m} \scal{u_j(-\kappa)}{u_j'(-\kappa)} \\
& = \int_{-2\pi}^{0} \di \kappa \, \sum_{i=1}^{m} \scal{u_i(\kappa)}{u_i'(\kappa)} = \int_{0}^{2\pi} \di \kappa \, \sum_{i=1}^{m} \scal{u_i(\kappa)}{u_i'(\kappa)}\,,
\end{align*}
where in the last equality we shifted the integration interval thanks to the $2\pi$-periodicity of the Bloch basis. As before, this allows to conclude that
\[ \frac{1}{2\pi \iu} \, \int_{S^1} \di k \, \sum_{i=1}^{2m} \scal{u_i(k)}{u_i'(k)} = \frac{1}{\pi \iu} \, \int_{S^1} \di k \, \sum_{i=1}^{m} \scal{u_i(k)}{u_i'(k)} \]
as wanted.
\end{proof}

\subsection{The $\Z_2$ invariant}

The above result finally puts in the position to set the following

\begin{definition}[The $\Z_2$ invariant]  \label{dfn:invariant}
Let $H$ be a finite-range Hamiltonian on $\ell^2(\Z) \otimes \C^N$ modeling a translation-invariant insulator, see Assumptions~\ref{hyp_main} and~\ref{finite_range}. Assume also that $H$ is either particle-hole symmetric or chiral symmetric, in the sense of Assumption~\ref{hyp_PHS} or~\ref{hyp_CS} respectively. Then the quantity 
\begin{equation}\label{eq:invariant}
    I\sub{PHC}(H) := \frac{1}{\pi \iu} \, \int_{S^1} \di k \, \sum_{i=1}^{m} \scal{u_i(k)}{u_i'(k)} \bmod 2 \quad \in \mathbb{Z}_2
\end{equation}
is a gauge invariant of the spectral projection of $H$ onto the negative energy eigenspace. Here $\set{u_i(k)}_{i\in{1,\ldots,N}}$ is any symmetric Bloch basis, in the sense of Definition~\ref{def:basis_comp_sym}.
\end{definition}

The above definition of our $\Z_2$ invariant $I\sub{PHC}$ is formulated in terms of a symmetric basis for the spectral projections $P_-(k)$ corresponding to negative eigenvalues of the fiber Hamiltonians; the symmetry dictates how the corresponding basis for $P_+(k) = \Id - P_-(k)$ should be defined. This could make the numerical computation of the invariant in specific models more accessible.

The next result states that $I\sub{PHC}(H)$ is also invariant under continuous deformations, i.e.\ homotopies, of the model, and therefore qualifies as a ``topological invariant''.

\begin{theorem}[Topological invariance of $I\sub{PHC}$]
Assume that $H_t$, $t \in [0,1]$, is a continuous family of bounded operators in $\ell^2(\Z) \otimes \C^N$ with the properties described in Definition~\ref{dfn:invariant}. Assume that the spectral gap of $H_t$ stays open for all $t \in [0,1]$. Then
\[  I\sub{PHC}(H_0) = I\sub{PHC}(H_1)\,. \]
\end{theorem}
\begin{proof}
The homotopy $t \mapsto H_t$ induces a corresponding homotopy $t \mapsto P_-(k;t)$ between the spectral projections of the fiber operators $H_t(k)$, which remains continuous due to the fact that the gap remains open. Since both $k \in S^1$ and $t \in [0,1]$ range over compact sets, by uniform continuity there exists $\delta > 0$ such that
\[ \sup_{k \in S^1} \norm{P_-(k;t) - P_-(k;s)} < 1 \quad \text{if} \quad |t-s| < \delta\,. \]
Let us use the short-hand notation $P_t(k) := P_-(k;t)$ and $P_s(k) := P_-(k;s)$ in the following. The above implies the existence of a \emph{Kato--Nagy unitary} (see Ref.~\onlinecite[Chapter I, Section 4.6]{Kato1995}) 
\[ U_{t,s}(k) := \left[ \Id - \left( P_t(k) - P_s(k) \right)^2 \right]^{-1/2} \, \left[ P_t(k) \, P_s(k) + \left(\Id - P_t(k)\right) \, \left(\Id-P_s(k)\right) \right] \]
such that
\[ P_t(k) \, U_{t,s}(k) = U_{t,s}(k) \, P_s(k)\,. \]
From the above explicit expression for the Kato--Nagy unitary, it is clear that it is  as regular in $k$ as the projections, jointly continuous in $t$ and $s$, and commutes with $\phs$ or $\cs$ if the models are particle-hole symmetric or chiral symmetric, respectively. 

Let us fix for example $s=0$, and let $\set{u_i(k;0)}_{i \in \set{1,\ldots,N}}$ be a symmetric Bloch basis associated to $P_-(k;0)$. By the above properties of the Kato--Nagy unitary, the collection of vectors
\[ u_i(k;t) := U_{t,0}(k) \, u_i(k;0)\,, \quad i \in \set{1,\ldots,N}\,, \]
defines a symmetric Bloch basis associated to $P_-(k;t)$, as long as $|t|< \delta$. Arguing as in the proof of Proposition~\ref{prop:even}, one can check that the Berry phase of the Bloch basis $\set{u_i(k;t)}_{i \in \set{1,\ldots,N}}$ has the same parity of the one of the Bloch basis $\set{u_i(k;0)}_{i \in \set{1,\ldots,N}}$, which defines the invariant $I\sub{PHC}(H_0)$. This implies that $I\sub{PHC}(H_t)$ is constant on the interval $[0,\delta)$. Partitioning the interval $[0,1]$ into a sequence of intervals of length $\delta/2$ and iterating this argument yields the conclusion of the proof.
\end{proof}

\begin{remark}
It is important to stress that, in the above statement, the symmetry (be it chiral or particle-hole) is required to be \emph{unbroken} along the whole homotopy $t \mapsto H_t$, on top of the assumption that the continuous deformation of $H_0$ to $H_1$ does not close the spectral gap. Indeed, one could connect also Hamiltonians which are in different topological phases (meaning that they have different values of $I\sub{PHC}$) without closing the gap, but breaking the symmetry along the deformation. The authors of Ref.~\onlinecite{ShapiroWeinstein2022}, for example, provide a deformation of (continuum models for) a ``topological'' chiral chain to a ``trivial'' one which keeps the spectral gap open, but do not claim that the deformation also leaves the chiral symmetry unbroken.
\end{remark}


\section{SSH model} \label{sec:SSH}

The Su--Schrieffer--Heeger (SSH) model, introduced in Ref.~\onlinecite{SSH1980} to model a molecular chain of polyacetylene, is often considered as a prototypical example of a chiral-symmetric topological insulator. The Hamiltonian $H\sub{SSH} \equiv H\sub{SSH}(\delta)$ acts on the Hilbert space $\Hi = \ell^2 (\Z) \otimes \C^2$ as
\[ \left(H\sub{SSH} \psi\right)_n = A_1 \, \psi_{n+1} + A_0 \, \psi_n + A_1^* \, \psi_{n-1}\,, \quad \psi \in \ell^2 (\Z) \otimes \C^2\,, \]
where
\[
A_1= \begin{bmatrix} 0 & 0 \\ 1 & 0 \end{bmatrix} , \quad A_0 \equiv A_0(\delta)= \begin{bmatrix} 0 & \delta \\ \delta & 0 \end{bmatrix} \,, \quad \text{with} \quad \delta \in \R\,.
\]
The SSH Hamiltonian thus fits Assumptions~\ref{hyp_main} and~\ref{finite_range}, provided we exclude some values of the parameter $\delta$ which lead to gapless spectrum. To this end, we apply Lemma~\ref{lem:decomposition_H} to obtain the fiber operators
\[
H\sub{SSH}(k) = A_0 + \eu^{-\iu k} \, A_1 + \eu^{\iu k} \, A_1^* = \begin{bmatrix}  0 & \delta+\eu^{\iu k} \\ \delta+\eu^{-\iu k} & 0 \end{bmatrix}
\]
with eigenvalues (\emph{Bloch bands})
\begin{equation} \label{eqn:ev_SSH}
E_{\pm}(k) := \pm \sqrt{(\delta +\cos k)^2 + (\sin k)^2}.
\end{equation}
The SSH model represents an insulator if and only if $E_{\pm}(k) \ne 0$ for all $k \in \R$, that is, if $\delta \notin\{-1, 1\}$.

Moreover, it is easily verified that this model verifies the CS hypothesis (or, more appropriately, a sublattice symmetry hypothesis), Assumption~\ref{hyp_CS}, with respect to the CS operator $\CS = \Id_{\ell^2(\Z)} \otimes \cs$ with $\cs = \sigma_3$, the third Pauli matrix, i.e.
\[
\cs \, \begin{bmatrix} x \\ y \end{bmatrix} = \begin{bmatrix} x \\ -y \end{bmatrix} \, .
\]
The SSH model is thus amenable to our analysis conducted in the previous Sections: as we will see, the parameter $\delta$ can be used to toggle topological and non-topological phases.

\subsection{The $\Z_2$ invariant for the SSH model}

In order to compute the invariant $I\sub{PHC}(\delta) \equiv I\sub{PHC}(H_{SSH}(\delta))$ in~\eqref{eq:invariant}, let us pick a smooth and periodic eigenvector for $H_{SSH}(k)$ with negative eigenvalue. To this end, let us denote $z(k) := \delta + \eu^{\iu k}$, so that in particular $E_\pm(k) = \pm |z(k)|$ and $z(k) \ne 0$ as long as the Hamiltonian is an insulator.
Solving the eigenvalue equation $H(k) \, u(k) = E_-(k) \, u(k)$ yields the normalized eigenvector
\[ u(k) = \frac{1}{\sqrt{2}} \, \begin{bmatrix} -1 \\ \sqrt{\dfrac{\overline{z(k)}}{z(k)}} \end{bmatrix} \]
from which we find, with a straightforward computation,
\[ \scal{u(k)}{u'(k)} = \frac{1}{4} \, \frac{\overline{z'(k)} \, z(k) -\overline{z(k)} \, z'(k)}{\overline{z(k)}z(k)} = - \frac{\iu}{2} \, \im \frac{z'(k)}{z(k)}\,.
\]
Therefore the $\Z_2$ invariant can be computed as
\begin{align*}
I\sub{PHC}(\delta) & = \frac{1}{\pi \iu} \, \int_{S^1} \di k \, \scal{u(k)}{u'(k)} \bmod 2 = - \frac{1}{2\pi} \, \int_{S^1} \di k \,\im \frac{z'(k)}{z(k)} \bmod 2 \\
& = \re \left[ - \frac{1}{2\pi\iu} \, \int_{S^1} \di k \, \frac{z'(k)}{z(k)} \right] \bmod 2\,.
\end{align*}
By the Cauchy integral formula, the integral above computes the winding number around the origin of the curve $\gamma$ in the complex plane obtained as the image of $k \mapsto z(k) = \delta + \eu^{\iu k}$, namely a circle of unit radius centered around $\delta \in \R \subset \C$. This winding number is an integer, and therefore the real part in right-hand side of the above identity is redundant: in particular, the curve $\gamma$ winds around the origin once if $\delta\in(-1,1)$ or zero times otherwise. (Notice that $\gamma$ passes through $0 \in \C$ if $\delta = \pm 1$, which are excluded values if we want the spectral gap to be open.) We conclude that
\[
I\sub{PCH}(\delta) = \begin{cases}
1 & \text{if } \delta\in(-1,1) \,, \\
0 & \text{if } \delta \in \R \setminus [-1,1]\,.
\end{cases}
\]

Incidentally, let us observe that the integer winding number of $k \mapsto z(k)$ coincides with the ``bulk'' topological invariant associated to $2$-band chiral chains in Refs.~\onlinecite{GrafShapiro2018, Shapiro2020}.

\subsection{Boundary modes for the SSH model}

When a topological insulator is cut along some edge, the \emph{bulk-boundary correspondence} predicts the appearance of new modes which fill the spectral gap of the bulk Hamiltonian and are spatially localized around the cut. We investigate this situation here for the SSH model, and verify that localized boundary modes to appear exactly for the values of the parameter $\delta$ for which $I\sub{PCH} \ne 0 \in \Z_2$. For a general proof of a bulk-edge correspondence in (disordered) chiral chains, we refer the reader to Ref.~\onlinecite{GrafShapiro2018}.

Let us define a truncated version of the SSH Hamiltonian as follows: it will be given by the operator $H\sub{SSH}^\sharp$ acting on $\Hi^\sharp := \ell^2(\N) \otimes \C^2$ as
\begin{gather*}
(H\sub{SSH}^\sharp \, \psi)_n := (H\sub{SSH} \, \psi)_n \quad \text{if } n > 0, \quad \text{and} \\
(H\sub{SSH}^\sharp \, \psi)_0 := A_1 \, \psi_1 + A_0 \, \psi_0 , \quad \psi \in \ell^2(\N) \otimes \C^2\,.
\end{gather*}
The truncated Hamiltonian imposes a Dirichlet condition on $H\sub{SSH}$ which sets $\psi_n=0$ for negative $n$. This boundary condition does not change the essential spectrum of $H\sub{SSH}$, which remains given by the two bands obtained as images of the functions defined in~\eqref{eqn:ev_SSH}, but pure point spectrum may appear in the bulk gap. We therefore look for zero-energy modes of $H\sub{SSH}^\sharp$, namely
for a vector $\psi = ([x_n , y_n]\su{T}) \in \Hi^\sharp$, $\psi \ne 0$, such that $H\sub{SSH}^\sharp \, \psi=0$. Explicitly, this reads
\begin{gather*}
\begin{bmatrix} 0 & 0 \\ 1 & 0 \end{bmatrix} \begin{bmatrix} x_{n+1} \\y_{n+1} \end{bmatrix} + \begin{bmatrix} 0& \delta \\ \delta & 0\end{bmatrix}\begin{bmatrix} x_n \\y_n \end{bmatrix} + \begin{bmatrix} 0 & 1 \\ 0 & 0 \end{bmatrix} \begin{bmatrix} x_{n-1} \\y_{n-1} \end{bmatrix} = \begin{bmatrix} 0 \\ 0 \end{bmatrix} \quad \text{if } n>0\,, \quad \text{and} \\[5pt]
\begin{bmatrix} 0 & 0 \\ 1 & 0 \end{bmatrix} \begin{bmatrix} x_{1} \\y_{1} \end{bmatrix} + \begin{bmatrix} 0& \delta \\ \delta & 0\end{bmatrix}\begin{bmatrix} x_0 \\ y_0 \end{bmatrix} = \begin{bmatrix} 0 \\ 0 \end{bmatrix} \,.
\end{gather*}
This yields the recursion rule
\[
\begin{cases}
\delta \, x_0 = -x_1\,, & \\
\delta \, y_0 = 0\,, & \\
x_{n+1}=-\delta \, x_{n} & \text{if } n>0\,, \\
\delta \, y_{n} = -y_{n-1} & \text{if } n>0\,. 
\end{cases}
\]
This obviously means that $x_n = (-\delta)^n \, x_0$ and $y_n \equiv 0$. In order for the resulting vector $\psi$ to be square-summable in the index $n \in \N$, we therefore have to require
\[
\norm{\psi}^2_{\Hi^\sharp} = \sum_{n \in \N} |x_n|^2 + |y_n|^2 = \sum_{n \in \N} |\delta|^{2n} < \infty \quad \Longleftrightarrow \quad |\delta|<1 \, .
\]
This range for $\delta$ coincides exactly with the values of the parameter in which the $\Z_2$ invariant $I\sub{PHC}(\delta)$ is non-trivial.


\section{Kitaev chain} \label{sec:Kitaev}

The Kitaev chain was introduced in Ref.~\onlinecite{Kitaev2001} as a model for a topological superconductor. In the notation introduced in Section~\ref{sec:BF} the Hamiltonian $H\sub{Kit} \equiv H\sub{Kit}(\mu,\delta)$ acts on the Hilbert space $\Hi=\ell^2(\Z) \otimes \C^2$ as
\[
(H\sub{Kit} \, \psi)_n = A_1 \, \psi_{n+1} + A_0 \, \psi_n + A_1^* \psi_{n-1}
\]
where
\[
A_1 \equiv A_1(\delta) = \begin{bmatrix} 0 & 1 + \delta \\ 1-\delta & 0 \end{bmatrix} \, , \quad A_0 \equiv A_0(\mu) = \begin{bmatrix} 0 & \mu \\ \mu & 0 \end{bmatrix} \, ,  \quad \mbox{with} \quad \delta , \mu \in \R.
\]
To check that this models an insulator, we compute the fiber operators as prescribed by Lemma~\ref{lem:decomposition_H} and obtain
\[ H\sub{Kit}(k)
= \begin{bmatrix} 0 & (1+\delta)\,\eu^{-\iu k} +\mu + (1-\delta) \, \eu^{\iu k} \\ (1-\delta) \, \eu^{-\iu k} + \mu +(1+\delta) \, \eu^{\iu k} & 0 \end{bmatrix}\,,
\]
which in turn yields the Bloch bands
\[
E_\pm(k) := \pm \sqrt{(\mu + 2 \cos k)^2 + (2 \delta \sin k)^2}\,.
\]
Therefore, the Kitaev Hamiltonian is gapped if and only if $E_{\pm}(k) \ne 0$ for every $k \in \R$: this leads to exclude the set of parameters $\mu,\delta$ given by
\[ \eta := \set{(\mu,\delta) \in \R^2 : |\mu| < 2 \text{ and } \delta = 0 \quad \text{or} \quad |\mu| = 2 \text{ and } \delta \in \R}\,. \]

If $(\mu, \delta) \in \R^2 \setminus \eta$, the Kitaev Hamiltonian $H\sub{Kit}$ thus verifies both Assumptions~\ref{hyp_main} and~\ref{finite_range}. Moreover, this model verifies the PHS hypothesis \ref{hyp_PHS} with respect to the operator $\PHS = \Id_{\ell^2(\Z)} \otimes \phs$ with
\[ \phs \, \begin{bmatrix} x \\ y \end{bmatrix} = \begin{bmatrix} \overline{x} \\ -\overline{y} \end{bmatrix}\,. \]
The Kitaev chain also satisfies the CS from the previous Section on the SSH model, but it is often considered paradigmatic for its anti-unitary particle-hole symmetry.

\subsection{The $\Z_2$ invariant for the Kitaev chain}

The invariant \[I\sub{PHC}(\mu,\delta) \equiv I\sub{PHC}(H\sub{Kit}(\mu,\delta)) \in \Z_2\] can be computed following the same lines of the argument for the SSH model, starting this time from the function 
\[ z(k) := (1+\delta) \, \eu^{-\iu k} + \mu + (1-\delta) \, \eu^{\iu k} = \mu + 2 \cos k - 2 \iu \delta \sin k, \quad k \in \R\,. \]
Its image $\gamma \subset \C$ describes an ellipse centered around $\mu \in \R$ with semi-axes $2$ and $2\delta$; its orientation is dictated by the sign of $\delta \in \R$. As before $I\sub{PHC}(\mu,\delta) \in \Z_2$ computes the parity of the winding number of this curve around the origin in $\C$, and therefore we conclude that
\begin{equation} \label{eqn:Z2Kitaev}
I\sub{PHC} (\mu,\delta) = \begin{cases} 1 & \text{if } |\mu|<2 \text{ and } \delta \ne 0\,, \\ 0 & \text{otherwise in } \R^2 \setminus \eta\,. \end{cases}
\end{equation}

\subsection{Boundary modes for the Kitaev chain}

As we did for the SSH model, we now pass to the investigation of zero-energy boundary modes for a truncated Kitaev chain. These modes are predicted to behave as Majorana particles, and their topological nature makes them amenable for applications in quantum computing as robust qubits. 

The truncated operator $H\sub{Kit}^\sharp$ acts on $\Hi^\sharp = \ell^2(\N) \otimes \C^2$ as
\begin{gather*}
(H\sub{Kit}^\sharp \, \psi)_n := (H\sub{Kit} \, \psi)_n \quad \text{if } n > 0, \quad \text{and} \\
(H\sub{Kit}^\sharp \, \psi)_0 := A_1 \, \psi_1 + A_0 \, \psi_0 , \quad \psi \in \ell^2(\N) \otimes \C^2\,.
\end{gather*}
This means that a vector $\psi = ([x_n , y_n]\su{T}) \in \Hi^\sharp$, $\psi \ne 0$, such that $H\sub{Kit}^\sharp \, \psi=0$ satisfies
\begin{gather*}
\begin{bmatrix} 0 & 1+\delta \\ 1-\delta & 0 \end{bmatrix} \begin{bmatrix} x_{n+1} \\y_{n+1} \end{bmatrix} + \begin{bmatrix} 0 & \mu \\ \mu & 0 \end{bmatrix} \begin{bmatrix} x_n \\y_n \end{bmatrix} + \begin{bmatrix} 0 & 1-\delta \\ 1+\delta & 0 \end{bmatrix} \begin{bmatrix} x_{n-1} \\y_{n-1} \end{bmatrix} = \begin{bmatrix} 0 \\ 0 \end{bmatrix} \quad \text{if } n>0\,, \quad \text{and} \\[5pt]
\begin{bmatrix} 0 & 1+\delta \\ 1-\delta & 0 \end{bmatrix} \begin{bmatrix} x_{1} \\y_{1} \end{bmatrix} + \begin{bmatrix} 0& \mu \\ \mu & 0\end{bmatrix}\begin{bmatrix} x_0 \\ y_0 \end{bmatrix} = \begin{bmatrix} 0 \\ 0 \end{bmatrix} \,,
\end{gather*}
leading this time to the decoupled recursion rules
\begin{equation} \label{eqn:recurrence_x}
\begin{cases}
(1-\delta) \, x_1 + \mu \, x_0 = 0\,, & \\
(1-\delta) \, x_{n+1} + \mu \, x_n + (1+\delta) \, x_{n-1}  = 0 & \text{if } n>0\,,
\end{cases}
\end{equation}
and
\begin{equation} \label{eqn:recurrence_y}
\begin{cases}
(1+\delta) \, y_1 + \mu \, y_0 = 0\,, & \\
(1+\delta) \, y_{n+1} + \mu \, y_n + (1-\delta) \, y_{n-1}  = 0 & \text{if } n>0\,.
\end{cases}
\end{equation}

\begin{remark} \label{rmk:x-y}
Notice that one is obtained from the other by the exchange of $\delta \leftrightarrow -\delta$: in particular, if the ``particle'' $\psi = ([x_n(\mu,\delta),0]\su{T}) \in \Hi^\sharp$ is a solution for the first recurrence relation, then the ``hole'' $\psi = ([0,x_n(\mu,-\delta)]\su{T}) \in \Hi^\sharp$ is a solution for the second recurrence equation, and viceversa. 
\end{remark}

\paragraph{\textsl{The case $\delta=\pm 1$}}

The situation is simpler when $\delta=\pm 1$: indeed, if $\delta = 1$ the relation~\eqref{eqn:recurrence_y} reduces to
\[ 
\begin{cases}
2 \, y_1 + \mu \, y_0 = 0\,, & \\
2 \, y_{n+1} + \mu \, y_n = 0 & \text{if } n>0\,,
\end{cases}
\]
which is of the form already encountered in the SSH model. In particular, the solution $y_n = (-\mu/2)^n \, y_0$ is square-summable if and only if $|\mu| < 2$. Notice that $I\sub{PHC} (\mu,1)$ is non-trivial exactly in this range of parameters. Similar considerations hold for~\eqref{eqn:recurrence_y} starting from $\delta = -1$.

\paragraph{\textsl{The characteristic equation}}

From now one we will suppose $\delta \ne \pm 1$. Moreover, in view of Remark~\ref{rmk:x-y}, it will be enough to restrict the range of parameters to $(\R^2 \setminus \eta) \cap \set{\delta \ge 0,\, \delta \ne 1}$ and solve the system~\eqref{eqn:recurrence_y}. Under these assumptions, the latter can be recast as
\begin{equation} \label{eqn:recurrence}
\begin{cases}
y_{n+1} + \dfrac{\mu}{1+\delta} \, y_n + \dfrac{1-\delta}{1+\delta} \, y_{n-1}  = 0 & \text{for } n \in \N\,, \\[5pt]
y_{-1} = 0, \quad y_0 \ne 0 &
\end{cases}
\end{equation}
(the condition on $y_0$ is to exclude the trivial solution $y_n \equiv 0$ and hence $\psi = 0 \in \Hi^\sharp$ which cannot define an eigenvector for $H\sub{Kit}^\sharp$). The theory of homogeneous linear difference equations with constant coefficients~\cite{Batchelder1927} prompts to consider the \emph{characteristic equation} 
\begin{equation} \label{eqn:characteristic}
\lambda^2 + \dfrac{\mu}{1+\delta} \, \lambda + \dfrac{1-\delta}{1+\delta} = 0\,.
\end{equation}
If $\lambda_+ \ne \lambda_-$ are the distinct roots of the above equation, then the general solution of~\eqref{eqn:recurrence} is given by
\[ y_n = a \, \lambda_+^n + b \, \lambda_-^n\,, \]
while if $\lambda_+ = \lambda_- = \lambda$ solutions are of the form
\[ y_n = (a + b \, n) \, \lambda^n\,. \]
In both cases, $a, b$ are constants chosen to fit the initial conditions for the recursion: in particular they cannot be simultaneously zero if we want a non-trivial solution. The above solutions can be easily verified by plugging in the recurrence relation an {\it ansatz} of the form $y_n = \lambda^n$. 

Let us consider first the case $\lambda_+=\lambda_-=\lambda$. In order to have that $\psi=([0,y_n]\su{T})$ is square-summable, we need to impose that
\[ \norm{\psi}^2_{\Hi^\sharp} = \sum_{n \in \N} |y_n|^2 = \sum_{n \in \N} |a+b\,n|^2 \, |\lambda|^{2n} < \infty\,. \]
The behaviour of the series on the right-hand side is dictated by its geometric part, and hence it converges if and only if $|\lambda|<1$. In the case $\lambda_+ \ne \lambda_-$, we appeal to the following
\begin{lemma}
Given $a,b \in \C$, the series
\[ \sum_{n \in \N} \big| a \, \lambda_+^n + b \, \lambda_-^n \big|^2 \] converges if and only if $|\lambda_+|<1$ and $|\lambda_-|<1$.
\end{lemma}
\begin{proof}
The statement is clearly true if either $\lambda_+=0$ or $\lambda_- = 0$, or if $a=0$ or $b=0$. Hence let us assume that $a \ne 0 \ne b$ and, without loss of generality, that $|\lambda_+| > |\lambda_-| > 0$. If $|\lambda_+| > 1$ then
\[
\big| a \, \lambda_+^n + b \, \lambda_-^n \big|^2 = |\lambda_+|^{2n} \, \left|a + b \, \left(\frac{\lambda_-}{\lambda_+}\right)^n \right|^2 \longrightarrow \infty  \quad \mbox{for } n \to \infty
\]
so the series cannot converge. Viceversa, if both $\lambda_+$ and $\lambda_-$ are smaller than 1 in absolute value, we can write
\[
\sum_{n \in \N} \left| a \, \lambda_+^n + b \, \lambda_-^n \right|^2 = \sum_{n \in \N} |a|^2 \, \left| \lambda_+ \right|^{2n} + |b|^2 \, \left| \lambda_- \right|^{2n} + 2\,\re \left[ \overline{a} \, b \, \left( \overline{\lambda_+} \, \lambda_- \right)^n \right]
\]
which is a sum of geometric series whose ratios are complex numbers with modulus smaller than 1, so they converge. 
\end{proof}

With the above considerations at hand, we are led to investigate the roots of the characteristic equation~\eqref{eqn:characteristic}, namely 
\[
\lambda_{\pm} := \frac{1}{2} \left[ - \frac{\mu}{1+\delta} \pm \sqrt{\frac{\mu^2 - 4 \, (1-\delta)\,(1+\delta)}{(1+\delta)^2}} \right] = \frac{1}{2} \left[ - \frac{\mu}{1+\delta} \pm \sqrt{\frac{\mu^2 - 4 + 4\,\delta^2}{(1+\delta)^2}} \right]
\]
and to study when they are both smaller than 1 in absolute value.

\paragraph{\textsl{Coinciding roots of the characteristic equation}}
If $\lambda_+ = \lambda_- = -\mu/2(1+\delta)$, then observing that we are restricting to $\delta \ge 0$ we have that
\[ \left|- \frac{\mu}{2\,(1+\delta)}\right| < 1 \quad \Longleftrightarrow \quad |\mu| < \min_{\delta \ge 0} |2 \, (1+\delta)| = 2\,. \]

\paragraph{\textsl{Complex roots of the characteristic equation}}
If $\mu^2 + 4 \delta^2 < 4$, the characteristic equation has two distinct complex roots
\[
\lambda_{\pm} = \frac{1}{2} \left[ - \frac{\mu}{1+\delta} \pm \iu \, \sqrt{\frac{4 - 4\,\delta^2 - \mu^2}{(1+\delta)^2}} \right]\,.
\]
Since $|\lambda|<1$ if and only if $|\lambda|^2 < 1$, we can bound
\[ \left| \lambda_+ \right|^2 = \left| \lambda_- \right|^2 = \frac{1}{4} \left[ \frac{\mu^2}{(1+\delta)^2} + \frac{4 - 4 \,\delta^2 - \mu^2}{(1+\delta)^2} \right] = \frac{1-\delta^2}{(1+\delta)^2} = 1 - \frac{2\,\delta}{1+\delta} < 1 \]
for all $\delta > 0$. If $\delta = 0$, the relation $\mu^2 + 4 \delta^2 < 4$ forces $|\mu| < 2$: but the region $\set{|\mu|<2 \text{ and } \delta = 0}$ in parameter space lies in $\eta$, and is therefore excluded from our analysis. We conclude that once again square-integrable solutions of the recurrence relation~\eqref{eqn:recurrence} exists if $\mu^2 + 4 \delta^2 < 4$ for all $\delta > 0$, that is, for $|\mu|<2$.

\paragraph{\textsl{Real roots of the characteristic equation}}
Finally, when $\mu^2+4\delta^2>4$ the characteristic equation has two distinct real roots
\[
\lambda_{\pm} = \frac{1}{2} \left[ - \frac{\mu}{1+\delta} \pm \sqrt{\frac{\mu^2 - 4 + 4\,\delta^2}{(1+\delta)^2}} \right]\,.
\]
We are led this time to solve the system
\[
\begin{cases}
\dfrac{\mu^2}{4(1+\delta)^2}+\dfrac{\mu^2+4\delta^2-4}{4\left(1+\delta\right)^2} -\dfrac{\mu}{1+\delta}\sqrt{\dfrac{\mu^2+4\delta^2-4}{4\left(1+\delta\right)^2}}<1\\[5pt]
\dfrac{\mu^2}{4(1+\delta)^2}+\dfrac{\mu^2+4\delta^2-4}{4\left(1+\delta\right)^2} +\dfrac{\mu}{1+\delta}\sqrt{\dfrac{\mu^2+4\delta^2-4}{4\left(1+\delta\right)^2}}<1
\end{cases} 
\]
which after a lengthy but straightforward computation is shown to be equivalent to $|\mu|<2$.

\paragraph{\textsl{Conclusions}}
In conclusion, if we combine all the previous results we obtain that, if $\delta \ge 0$ (and, in view of Remark~\ref{rmk:x-y}, also if $\delta\le0$), there is a zero-energy boundary state for the truncated Kitaev chain if and only if $|\mu|<2$. Comparing with~\eqref{eqn:Z2Kitaev}, we conclude that as in the SSH model the presence of boundary states occurs exactly for those values of the physical parameters $\mu,\delta$ for which the $\Z_2$ invariant is non-trivial.


\begin{acknowledgments}
D.~M.~wishes to thank the organizers of the workshop ``Learning from Insulators: New Trends in the Study of Conductivity of Metals'' (Giuseppe De Nittis, Max Lein, Constanza Rojas-Molina, and Marcello Seri) for the invitation to participate in the event and their committment to its organization even in the most trying times of the COVID-19 pandemic. The authors also gratefully acknowledge financial support from Sapienza Universit\`{a} di Roma within Progetto di Ricerca di Ateneo 2020 (grant no.\ RM120172AE419BE1) and 2021 (grant no.\ RM12117A86FB96EE).

This work has been carried out under the auspices of the GNFM--INdAM (Gruppo Nazionale per la Fisica Matematica -- Istituto Nazionale di Alta Matematica), and within the framework of the activities for PNRR MUR Project PE0000023-NQSTI.
\end{acknowledgments}

\section*{Data Availability Statement}

Data sharing is not applicable to this article as no new data were created or analyzed in this study.


\appendix

\section{Winding numbers for unitary matrices} \label{sec:appendix}

This Appendix is devoted to give some insights into the proof of Theorem~\ref{thm:p1(UN)}; more details can be found e.g.\ in Ref.~\onlinecite[Section III]{MonacoRoussigne2022}. The statement that $U(N)$-valued maps on $S^1$ are characterized by their winding number up to homotopy can be essentially reduced to the analogous fact for $N=1$, namely that maps $S^1 \to U(1) \simeq S^1$ are described by the way they wind around the origin in the complex plane. The reduction step is based on the following

\begin{proposition}\label{prop:derivative_determinant}
If $U \colon \R \to U(N)$ is at least $C^1$, then
\[
\frac{\partial_k \, \det U(k)}{\det U(k)} =  \tr \big(U(k)^* \, U'(k)\big) \,.
\]
\end{proposition}
\begin{proof}
Since $U(k)$ is unitary we have $U(k) U(k)^* = \Id$, and using the fact that $\det ( \Id + H ) = 1 + \tr(H) + \mathcal{O}(\norm{H}^2)$ we can compute
\begin{align*}
\partial_k \det U(k) & = \lim_{h\to 0} \frac{\det U(k+h) -\det U(k)}{h} = \det U(k) \cdot \lim_{h\to 0} \frac{\det\big(\Id - U(k)^* \, U(k) + U(k)^* \, U(k+h)\big) - 1}{h} \\
& = \det U(k) \cdot \lim_{h\to 0} \frac{\tr \big(U(k)^* \, U(k+h) - U(k)^* \, U(k) \big)}{h} = \det U(k) \cdot \tr \left\{ U(k)^* \cdot \left[ \lim_{h \to 0} \frac{U(k+h) - U(k)}{h} \right] \right\} \\
& = \det U(k) \cdot \tr \big( U(k)^* \, U'(k) \big)\,. \qedhere
\end{align*}
\end{proof}

The above result immediately gives that $\wn([U])$ defined in~\eqref{eqn:wn} can be computed as 
\[ \wn([U]) = \frac{1}{2\pi\iu} \, \int_{S^1} \di k \, \frac{\partial_k \, \det U(k)}{\det U(k)}\,. \]
We thus shift our focus on the function $u := \det U \colon S^1 \to U(1) \simeq S^1$.

\begin{proposition}\label{cor:top_group}
Let $f,g \colon S^1 \to S^1$ be two differentiable maps. The integral
\[ w(f) := \frac{1}{2\pi\iu} \int_{S^1} \di k \, \frac{f'(k)}{f(k)} \]
is integer-valued, and depends only on the homotopy class of $f$: it is called the \emph{winding number} of $f$. Moreover, the winding number of the product is the sum of the two winding numbers:
\[
w(f \cdot g) = w(f) + w(g).
\]
\end{proposition}
\begin{proof}
The image of $f \colon S^1 \to S^1$ can be seen as a rectifiable and closed curve $\gamma \subset \C$. By the Cauchy integral formula
\[
\int_{S^1} \di k \, \frac{f'(k)}{f(k)} = \int_\gamma \di z \, \frac{1}{z} = 2 \pi \iu \, \text{Ind}_\gamma(0) 
\]
where $\text{Ind}_\gamma(0)$ computes the winding number of the curve $\gamma$ around the origin in $\C$. As is well known, this quantity is integer-valued.

If $f_s \colon S^1 \to S^1$, $s \in [0,1]$, is a smooth homotopy between two maps $f_0$ and $f_1$, then
\[
\frac{\di}{\di s} \int_{S^1} \di k \, \frac{f_s'(k)}{f_s(k)} 
 = \int_{S^1} \di k \, \frac{\partial_s \partial_k f_s(k) \, f_s(k) - \partial_k f_s(k) \, \partial_s f_s(k)}{f_s(k)^2} 
= \int_{S^1} \di k \, \partial_k \, \left( \frac{\partial_s f_s(k)}{f_s(k)} \right) = 0
\] 
and therefore $w(f)$ is constant along the homotopy class of $f$.

Finally, to show the additive property of the winding number, it suffices to compute
\[
w(f \cdot g) 
= \frac{1}{2\pi \iu}\,\int_{S^1} \di k \, \frac{\partial_k (fg)(k)}{f(k) g(k)} 
= \frac{1}{2\pi \iu}\,\int_{S^1} \di k \, \frac{\partial_k f(k)}{f(k)} + \frac{1}{2\pi i}\,\int_{S^1} \di k \, \frac{\partial_k g(k)}{g(k)} dk 
= w(f)+w(g)\,. \qedhere
\]
\end{proof}

\begin{remark} \label{rmk:wind_inv}
Let $\iota \colon S^1 \to S^1$ be the involution $k \mapsto -k$. For $f \colon S^1 \to S^1$, we then have
\[ w(f \circ \iota) = - w(f)\,. \]
Indeed, we can compute $(f \circ \iota)'(k) = - f'(-k)$ and therefore
\[
2 \pi \iu \, w(f \circ \iota) 
= - \int_{0}^{2\pi} \di k \, \frac{f'(-k)}{f(-k)} = \int_{0}^{-2\pi} \di \kappa \, \frac{f'(\kappa)}{f(\kappa)} 
= - \int_{-2\pi}^{0} \di \kappa \, \frac{f'(\kappa)}{f(\kappa)} = - \int_{0}^{2\pi} \di \kappa \, \frac{f'(\kappa)}{f(\kappa)} = - 2\pi\iu \, w(f)
\]
where in the second equality we used the substitution $\kappa = \iota(k)=-k$ (from which $\di \kappa = - \di k$) and in the second-to-last equality we shifted the integration interval by periodicity of $f$.
\end{remark}

\begin{proposition} \label{prop:compute_winding_number}
The map
\[
\pi_1 \left(S^1\right) \to \Z, \quad [f] \mapsto \frac{1}{2\pi \iu} \, \int_{S^1} \di k \,\frac{f'(k)}{f(k)}
\]
is a bijection. 
\end{proposition}
\begin{proof}
See e.g.\ Ref.~\onlinecite[Chapter 1, Section 5]{May1999} or Ref.~\onlinecite[pages 7-8]{MonacoRoussigne2022}.
\end{proof}



\end{document}